\documentclass[12pt]{iopart}
\usepackage[totalheight=11in, totalwidth=8.5in, top=1in, bottom=1in, left=1in, right=1in]{geometry}
\usepackage{amsfonts}
\usepackage{wasysym}
\usepackage{graphicx}
\usepackage{comment}
\usepackage{tensor}
\usepackage{cancel}
\usepackage[svgnames]{xcolor}

\usepackage[square,numbers,sort&compress]{natbib}
\usepackage[colorlinks = true, linkcolor = DarkBlue, citecolor = DarkBlue, urlcolor = DarkBlue, bookmarks = false]{hyperref}

\def\filetype{pdf}
\def\path{}

\begin{document}

%---TITLE PAGE------------------------------------------------------------------

\title{Dynamical wormholes}
\author{Ben Kain}
\address{Department of Physics\\ College of the Holy Cross\\ Worcester, Massachusetts 01610, USA}

\begin{abstract}
\noindent
We numerically investigate the dynamical evolution of spherically symmetric charge free wormholes. We concentrate on two specific examples, both of which exhibit wormhole expansion and wormhole collapse:~the Ellis-Bronnikov wormhole, which is sourced by a real massless ghost scalar field, and the quantum corrected Schwarzschild black hole in semiclassical gravity (which has a wormhole structure and is not a true black hole), which is sourced by a renormalized energy-momentum tensor. Despite their very different sources, we demonstrate that the dynamics of these two wormholes are remarkably similar. Our analysis focuses on diagrams for the areal radius and components of the energy-momentum tensor. This work also serves as a review, offering a detailed description of how to perform a spherically symmetric dynamical evolution using double null coordinates as well as a review of the static solutions for our two examples.
\end{abstract} 

%\maketitle
\tableofcontents

%==================================================

\section{Introduction}

Wormholes are hypothetical tunnels connecting two distinct regions of spacetime \cite{VisserBook}. They arise as solutions to the gravitational field equations and have attracted significant interest since the seminal work of Morris and Thorne \cite{Morris:1988cz}, who showed that static spherically symmetric wormholes could be traversable when sourced with matter that violates the null energy condition. In this work, we investigate the dynamical evolution of spherically symmetric charge free wormholes. By dynamical evolution, we mean using a static wormhole solution as an initial configuration and numerically evolving the system in time. 

There are important reasons to study the dynamical evolution of static solutions. Static solutions can be unstable with respect to time dependent perturbations and a dynamical evolution can test the nonlinear stability. If a static solution is unstable, the dynamical evolution can reveal the system's final state. Unstable static solutions are unlikely to form naturally and cannot exist as the final state of an evolution. One may think a particular static wormhole solution is traversable, but if it unstable it could collapse and form horizons, rendering it temporarily traversable at best. A dynamical evolution may therefore be necessary to determine if a wormhole solution is genuinely traversable. 

The dynamical evolution of a static solution is also interesting in its own right. We will focus on unstable static solutions which exhibit expansion or collapse and we will investigate how the evolution can be understood from diagrams of the areal radius and components of the energy-momentum tensor. In doing this, we concentrate on two specific examples. The first example is the Ellis-Bronnikov wormhole \cite{Ellis:1973yv, Bronnikov:1973fh}, which has been extensively studied (for a sampling of work, see \cite{Gonzalez:2009hn, Bronnikov:2012ch, Tsukamoto:2016qro, Nandi:2016uzg, Dzhunushaliev:2017syc, Konoplya:2018ala, Cremona:2018wkj, Blazquez-Salcedo:2018ipc, Harko:2018gxr, Cremona:2019wiy, Bronnikov:2021ods, Yang:2021diz, Bronnikov:2021liv, Sokoliuk:2022xcf, Blazquez-Salcedo:2022kaw, Filho:2023yly}). This wormhole is sourced by a real massless ghost scalar field, where a ghost field is one which has the opposite sign in front of its kinetic term in the Lagrangian compared to a regular field. The physicality of a ghost field is debatable, but an advantage of the Ellis-Bronnikov wormhole is its simplicity, as the wormhole solution can be found analytically. The study of dynamically evolved wormholes was initiated by Shinkai and Hayward when they dynamically evolved the Ellis-Bronnikov wormhole \cite{Shinkai:2002gv}, which has since been dynamically evolved by a number of groups using various numerical methods \cite{Gonzalez:2008xk, Doroshkevich:2008xm, Shinkai:2015aqa, Shinkai:2017xkx, Novikov:2019rus, Novikov:2020xwr, Calhoun:2022xrw, Xu:2025jad}.

Additional examples of dynamically evolved wormholes are scarce, underscoring that the dynamical evolution of wormholes remains an underexplored area of research. The Einstein-Dirac-Maxwell wormhole \cite{Blazquez-Salcedo:2020czn, Konoplya:2021hsm, Kain:2023pvp} has been dynamically evolved \cite{Kain:2023ore, Kain:2023ann}, but this system is charged and the evolution did not exhibit expansion or collapse, and the quantum corrected Schwarzschild black hole \cite{Fabbri:2005zn, Ho:2017joh, Ho:2017vgi, Arrechea:2019jgx} has been dynamically evolved \cite{Kain:2025lxd}, which did exhibit expansion and collapse. As our second example, we present the dynamical evolution of the quantum corrected Schwarzschild black hole. The static wormhole solution is found in semiclassical gravity \cite{Birrell:1982ix}, where the semiclassical correction is taken to be a renormalized energy-momentum tensor in the Polyakov approximation \cite{Davies:1976ei, davies, Polyakov:1981rd}. The static solution asymptotically matches the classical Schwarzschild solution, but the classical horizon has disappeared and is replaced with a wormhole structure.

The Ellis-Bronnikov wormhole is sourced by a ghost field and the quantum corrected Schwarzschild black hole is sourced by a renormalized energy-momentum tensor. Despite these very  different sources, their respective simulations are remarkably similar, as we will show. This leads us to speculate that these results are generic. If this is the case, it lends support to studying the Ellis-Bronnikov wormhole, notwithstanding its use of a ghost field, because of its particularly simple analytical static solutions.

This article is partly a review. We give a detailed description of how to perform a dynamical evolution in spherical symmetry using double null coordinates. We also review the static solutions for Ellis-Bronnikov wormholes and quantum corrected Schwarzschild black holes. We hope by reviewing these ideas that we are encouraging further study of dynamical wormholes. Additionally, we present several novel results, most notably in our analysis of the energy-momentum tensor for Ellis-Bronnikov wormholes and in our comparison of the two systems.

A note on our use of the term wormhole: Throughout this work, we shall use the term wormhole to describe a system whose areal radius has a nonzero minimum. This condition signifies the existence of a wormhole throat and such a system is more properly labeled as having a wormhole structure. The term wormhole typically implies well-behaved geometry on both sides of the throat, which will not always be the case in the examples we study. Nonetheless, for convenience, we will refer to such systems as wormholes.

In section \ref{sec:dynamical framework}, we review our dynamical evolution framework. In sections \ref{sec:EB} and \ref{sec:QCSBH}, we study Ellis-Bronnikov wormholes and quantum corrected Schwarzschild black holes, respectively. In both cases, we review the static solutions, give details on how to dynamically evolve the static solutions, and analyze the simulations. For the quantum corrected Schwarzschild black hole, we also discuss semiclassical gravity and the renormalized energy-momentum tensor. In section \ref{sec:conclusion}, we compare  results for our two examples and conclude. We set $c = G = \hbar = 1$ throughout.

%====================================================================

\section{Dynamical evolution framework}
\label{sec:dynamical framework}

%--------------------------------------------------------------------

\subsection{Double null coordinates}
\label{sec:double null}

We work in spherical symmetry and use double null coordinates. We write the metric as
\begin{equation} \label{metric}
ds^2 = -e^{2\sigma(u,v)}dudv + r^2(u,v) d\Omega^2,
\end{equation}
where $u$ is the outgoing null coordinate, $v$ is the ingoing null coordinate, $\sigma(u,v)$ and $r(u,v)$ parameterize the metric, where $r$ is the areal radius, and $d\Omega^2 = d\theta^2 + \sin^2\theta \, d\phi^2$. Only a few different coordinate systems have been used in the dynamical evolution of wormholes, with double null coordinates being the most common choice. An advantage of using double null coordinates is that large portions of the spacetime can be computed and then displayed in a single diagram. Further, if the wormhole collapses and black holes form, the spacetime behind the horizons can be straightforwardly computed. For the specific case of the quantum corrected Schwarzschild black hole, double null coordinates are well-suited for the inclusion of the renormalized energy-momentum tensor.

The Einstein field equations are
\begin{equation} \label{EFE}
R_{\mu\nu} - \frac{1}{2} g_{\mu\nu} R = 8\pi T_{\mu\nu},
\end{equation}
where $R_{\mu\nu}$ is the Ricci tensor, $R$ is the Ricci scalar, and $T_{\mu\nu}$ is the energy-momentum tensor. For the metric in (\ref{metric}), the $uu$ and $vv$ components give the constraint equations
\begin{eqnarray} \label{constraint eqs}
r_{,uu} 
&= 2 \sigma_{,u} r_{,u} - 4\pi r T_{uu} 
\nonumber \\
r_{,vv} 
&= 2 \sigma_{,v} r_{,v} - 4\pi r T_{vv},
\end{eqnarray}
where a subscripted comma denotes a partial derivative, and the $uv$ and $\theta\theta$ components lead to the evolution equations
\begin{eqnarray} \label{sigma r evo}
\sigma_{,uv} 
&= \frac{1}{4r^2} \left(4 r_{,u} r_{,v} + e^{2\sigma} 
- 16\pi r^2 T_{uv} -8\pi e^{2\sigma} T_{\theta\theta} \right)
\nonumber \\
r_{,uv} 
&= - \frac{1}{4r} \left(4 r_{,u} r_{,v} + e^{2\sigma} 
- 16\pi r^2  T_{uv} \right).
\end{eqnarray}
When we discuss the Ellis-Bronnikov wormhole and the quantum corrected Schwarzschild black hole, we will present the components of the energy-momentum tensor that are specific to these wormholes.

The Misner-Sharp mass function, $m(u,v)$, gives the total mass inside a radius $r(u,v)$. It is defined by
\begin{equation} \label{Misner-Sharp def}
1 - \frac{2m}{r} = g^{\mu\nu} r_{,\mu} r_{,\nu}.
\end{equation}
Using the metric in (\ref{metric}), the mass function is given by
\begin{equation} \label{mass function}
m = \frac{r}{2} \left( 1 +  4e^{-2\sigma} r_{,u} r_{,v} \right).
\end{equation}
When evolving wormholes we will find apparent horizons, which are defined by $r_{,u} = 0$ or $r_{,v} = 0$. Along an apparent horizon, the mass is given by $m = r/2$. As we'll see, when a wormhole expands, the apparent horizon evolves along increasing radii and the mass inside the horizon increases. Conversely, when a wormhole collapses, the apparent horizon evolves along decreasing radii and the mass decreases. 

%--------------------------------------------------------------------

\subsection{Raychaudhuri equation}
\label{sec:Raychaudhuri}

The Raychaudhuri equation describes the evolution of a congruence, which is a bundle of curves such that every point in a region of spacetime lies on only one curve. In particular, we will be interested in the Raychaudhuri equation for a congruence of null geodesics, which is \cite{Carroll:2004st}
\begin{equation}
k^\mu \partial_\mu \theta = - \frac{1}{2}\theta^2 - \sigma_{\mu\nu} \sigma^{\mu\nu} + \omega_{\mu\nu}\omega^{\mu\nu} - R_{\mu\nu} k^\mu k^\nu,
\end{equation} 
where 
\begin{equation} \label{expansion def}
\theta = \nabla_\mu k^\mu
\end{equation}
is the expansion, which describes the fractional rate of change of the cross sectional area of the congruence, and $k^\mu$ is a null vector field tangent to the geodesics. $\theta_{\mu\nu}$ and $\omega_{\mu\nu}$ are the shear and rotation tensors, both of which vanish in spherical symmetry. Using the Einstein field equation in (\ref{EFE}), we have $R_{\mu\nu} k^\mu k^\nu = 8\pi T_{\mu\nu} k^\mu k^\nu$, since $k^\mu$ is null. The Raychaudhuri equation reduces to
\begin{equation} \label{Ray eq}
k^\mu \partial_\mu \theta = - \frac{1}{2}\theta^2 - 8\pi T_{\mu\nu} k^\mu k^\nu.
\end{equation}

Since $k^\mu$ is null and tangent to geodesics, it satisfies
\begin{equation}
k^\mu k_\mu = 0,
\qquad
k^\mu \nabla_\mu k^\nu = 0.
\end{equation}
In double null coordinates, with the metric in (\ref{metric}), solutions to these equations are
\begin{equation}
k^\mu_{(u)} = e^{-2\sigma}(1,0,0,0),
\qquad
k^\mu_{(v)} = e^{-2\sigma}(0,1,0,0).
\end{equation}
Each of these vectors is defined up to an overall arbitrary constant, which we have set to 1 by convention. With these $k^\mu$, we have two forms for the Raychaudhuri equation,
\begin{equation} \label{Ray eq double null}
\theta_{,u}^{(u)} = - \frac{1}{2} e^{2\sigma} \theta^2_{(u)}  - 8\pi e^{-2\sigma} T_{uu},
\qquad
\theta_{,v}^{(v)} = - \frac{1}{2} e^{2\sigma} \theta^2_{(v)}  - 8\pi e^{-2\sigma} T_{vv}.
\end{equation}
where, from (\ref{expansion def}),
\begin{equation} \label{theta double null}
\theta_{(u)} = e^{-2\sigma} \frac{2r_{,u}}{r},
\qquad
\theta_{(v)} = e^{-2\sigma} \frac{2r_{,v}}{r}.
\end{equation}
If we plug these forms for the expansion into (\ref{Ray eq double null}), we find that (\ref{Ray eq double null}) reproduces the constraint equations in (\ref{constraint eqs}), which shows that in spherical symmetry the Einstein field equations contain the Raychaudhuri equation.

The Raychaudhuri equation plays a central role in the study of wormholes. For example, it can be used to derive that the existence of a wormhole implies violation of the null energy condition \cite{Morris:1988cz}. By definition, a wormhole throat is located where the cross sectional area of the areal radius has a minimum, which occurs when $\theta = 0$ and $k^\mu \partial_\mu \theta \geq 0$ \cite{VisserBook}. It follows immediately from the Raychaudhuri equation in (\ref{Ray eq}) that the presence of a wormhole throat requires
\begin{equation}
T_{\mu\nu} k^\mu k^\nu \leq 0.
\end{equation}
The null energy condition is violated if $T_{\mu\nu} k^\mu k^\nu < 0$. The Raychaudhuri equation can also be used to derive that the existence of a wormhole requires violation of the averaged null energy condition, which is a stronger requirement \cite{VisserBook}, but we will only make use of the null energy condition.

In double null coordinates, the null energy condition is violated if
\begin{equation}
e^{-4\sigma} T_{uu} < 0
\quad \textrm{or} \quad
e^{-4\sigma} T_{vv} < 0.
\end{equation}
This reduces to $T_{uu} < 0$ or $T_{vv} < 0$. However, we should keep in mind that the null energy along null geodesics contains the factor of $e^{-4\sigma}$. Intuitively, we might expect that an expanding wormhole, one for which the wormhole throat radius is increasing, would have negative null energy that decreases, evolving toward being more negative. Conversely, a collapsing wormhole, one for which the radius decreases, would have negative null energy that increases, evolving toward zero. We will find this to generally be the case for $e^{-4\sigma}T_{uu}$ and $e^{-4\sigma}T_{vv}$, but not necessarily for $T_{uu}$ and $T_{vv}$.

%--------------------------------------------------------------------

\subsection{Numerical methods}
\label{sec:numerical methods}

To dynamically evolve a wormhole, we numerically solve some subset of the complete list of equations. We need only solve a subset since the equations are not all independent. Our numerical scheme, which we review below, is designed to solve evolution equations. These include the equations in (\ref{sigma r evo}) for the metric fields, but not the constraint equations in (\ref{constraint eqs}). If there is any matter in the system, there will be additional evolution equations which describe the matter. We use the constraint equations in (\ref{constraint eqs}) for computing initial data and for code testing.

\begin{figure} 
\centering
\includegraphics[width=5.5in]{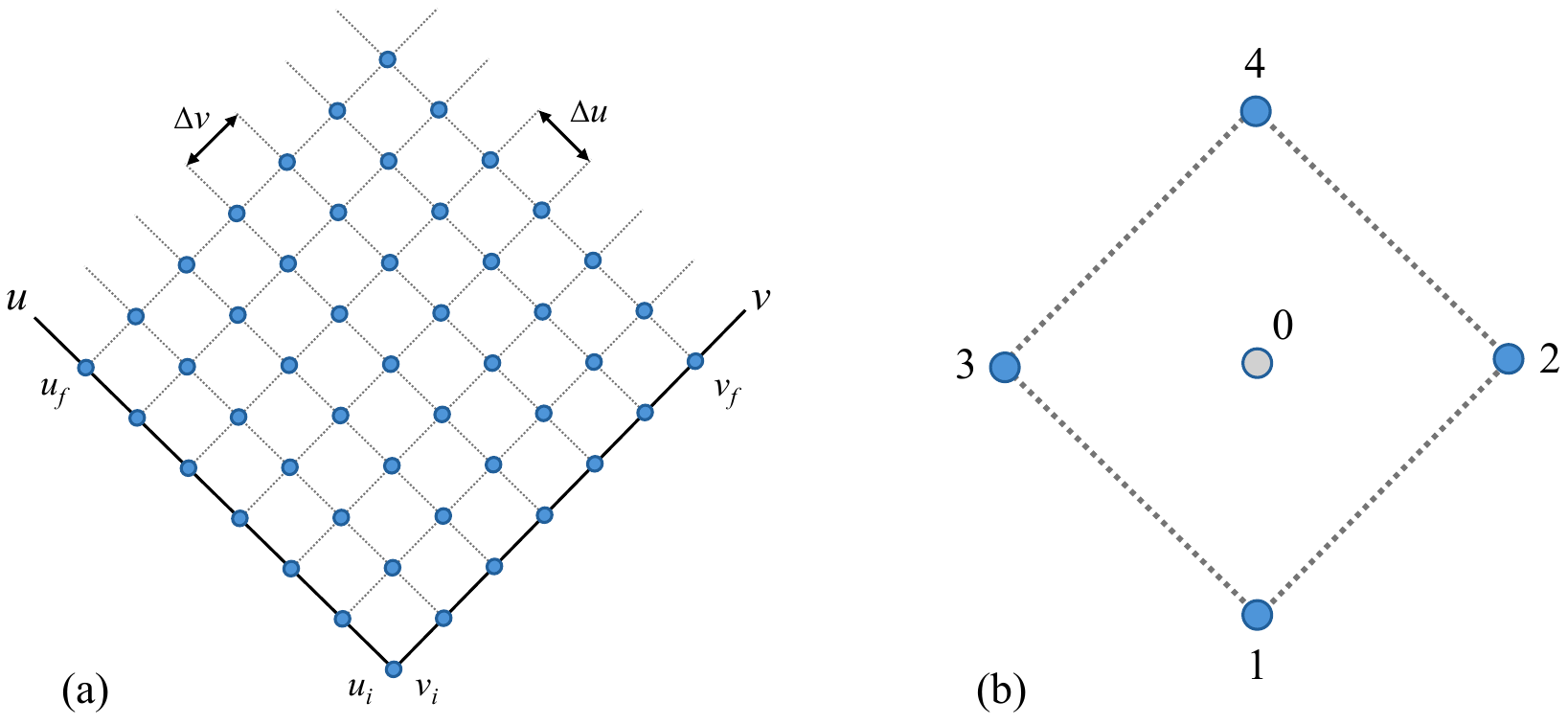} 
\caption{(a) A two dimensional discrete grid of $(u,v)$ values is used as our computational domain, here illustrated as a $7\times 7$ grid (in practice, each dimension will typically have thousands of grid points). The grid points have uniform spacing $\Delta u$ and $\Delta v$ and take values in the ranges $u_i \leq u \leq u_f$ and $v_i \leq v \leq v_f$. (b) A single grid cell, where points 1, 2, 3, and 4 are points on the grid, while point 0 is an intermediate point and is not on the grid. In general, we will know field values at points 1, 2, and 3 and will compute field values at point 4.}
\label{fig:grid cell}
\end{figure}

We use a two dimensional discrete grid of $(u,v)$ values as our computational domain. The grid points have uniform spacing $\Delta u$ and $\Delta v$ and take values in the ranges $u_i \leq u \leq u_f$ and $v_i \leq v \leq v_f$. $u = u_i$ and $v = v_i$ are the initial hypersurfaces. When plotting results, we set $u_i = v_i = 0$, but we keep $u_i$ and $v_i$ arbitrary in equations for completeness. We show a diagram for the grid in figure \ref{fig:grid cell}(a).

In figure \ref{fig:grid cell}(b), we show a single grid cell. In general, we will know the values of all fields at the grid points labeled 1, 2, and 3 and will numerically solve for field values at the grid point labeled 4. To do this, we used a standard predictor-corrector scheme, which is second order accurate, to numerically solve the evolution equations. This scheme makes use of the intermediate point labeled 0, which is not a point on our grid.

We now briefly review the numerical scheme (for additional details, see, for example, \cite{Eilon:2015axa}). The system of evolution equations takes the form
\begin{equation}
f_{,uv} = F(f, f_{,u}, f_{,v}),
\end{equation}
where $f$ is a vector of fields to solve for and $F$ is a vector of specified functions. Note that the highest derivative in $F$ is first order. This will always be the case. We will be evaluating these quantities at the labeled points in the unit cell shown in figure \ref{fig:grid cell}(b):~Let $f^{(i)}_{,uv}$, $f^{(i)}$, $f^{(i)}_{,u}$, $f^{(i)}_{,v}$ be the quantities evaluated at points $i = 0,1,\ldots,4$. As mentioned, we know the values at points 1, 2, and 3 and want to compute the values at point 4.

In the first predictor step, we compute fields at point 0 using
\begin{equation}
f^{(0)} = \frac{f^{(2)} + f^{(3)}}{2},
\qquad
\tilde{f}_{,u}^{(0)} = \frac{f^{(3)} - f^{(1)}}{\Delta u},
\qquad
\tilde{f}_{,v}^{(0)} = \frac{f^{(2)} - f^{(1)}}{\Delta v}.
\end{equation}
The last two formulas are finite differences which are only first order accurate, which is why we have included tildes. This is the predictor step, which will be corrected with a corrector step, yielding a fully second order accurate scheme. These quantities are then used in $F$ to compute
\begin{equation}
\tilde{f}_{,uv}^{(0)} = F(f^{(0)}, \tilde{f}_{,u}^{(0)}, \tilde{f}_{,v}^{(0)}),
\end{equation}
which is used in the formula
\begin{equation}
\tilde{f}^{(4)} = -f^{(1)} + f^{(2)} + f^{(3)} + \tilde{f}_{,uv}^{(0)} \Delta u \Delta v.
\end{equation}
The use of the predictor has led to a linear equation for $\tilde{f}^{(4)}$, but one that is only first order accurate, hence the tilde.

In the second corrector step, we compute
\begin{equation} \label{xU0 xV0}
f_{,u}^{(0)} = \frac{f^{(3)} - f^{(1)} + \tilde{f}^{(4)} - f^{(2)}}{2\Delta u},
\qquad
f_{,v}^{(0)} = \frac{f^{(2)} - f^{(1)} + \tilde{f}^{(4)} - f^{(3)}}{2\Delta v}.
\end{equation}
These quantities are then used in $F$ to compute
\begin{equation}
f_{uv}^{(0)} = F(f^{(0)}, f_{,u}^{(0)}, f_{,v}^{(0)}),
\end{equation}
which, as before, is used in the formula
\begin{equation}
f^{(4)} = -f^{(1)} + f^{(2)} + f^{(3)} + f_{,uv}^{(0)} \Delta u \Delta v.
\end{equation}
This result is second order accurate.

Our code begins by solving for the fields at grid point $(u_i + \Delta u, v_i + \Delta v)$ and then proceeds to solve for the fields at each grid point along the ``row" $u = u_i + \Delta u$ before proceeding to the subsequent row, $u = u_i + 2\Delta u$, and so on. In this way, our code computes the field values at each grid point in the computational domain. Of course, for this to be possible we must supply field values on the initial hypersurfaces $u = u_i$ and $v = v_i$. As we explain in the next subsection, we place a  static wormhole solution on the initial hypersurfaces. Our code therefore determines the dynamical evolution of a static solution subject to perturbations. 

We consider two types of perturbations. First, all code has an inherent discretization error which acts as a perturbation. By decreasing the grid spacing $\Delta u$ and $\Delta v$, we can decrease the strength of this perturbation, but it will always be present. This type of perturbation can be used to check the stability of static solutions, as we will see. In the presence of an explicit perturbation, however, a perturbation from discretization error should be negligible. As the second type of perturbation, we will introduce a matter field to be used as an explicit perturbation. When we discus the Ellis-Bronnikov wormhole and the quantum corrected Schwarzschild black hole, we give details for the matter field.

%--------------------------------------------------------------------

\subsection{Initial data}
\label{sec:initial data}

To dynamically evolve the system, we need the value of each field on the initial hypersurfaces. We refer to this set of values as initial data. For initial data, we use a static solution. More specifically, we will use a solution to the gravitational field equations that follow from a metric taking the form
\begin{equation} \label{static metric}
ds^2 = -\alpha^2(x) dt^2 + a^2(x) dx^2 + r^2(x) d\Omega^2,
\end{equation} 
where $t$ is the time coordinate, $x$ is the radial coordinate, $\alpha(x)$, $a(x)$, and $r(x)$ parameterize the metric, where $r$ is the areal radius. Since this metric is time independent and spherically symmetric, it describes a static spacetime. We would like the metric to be capable of describing a wormhole. We therefore allow the radial coordinate to take values in the range $-\infty < x < \infty$. 

Once we find a solution, we will have to transform it to double null coordinates to use it as initial data. We begin by introducing a new radial coordinate, $y$, defined by
\begin{equation} \label{dy dx}
\alpha \, dy = a \, dx,
\end{equation}
so that the metric becomes 
\begin{equation}
ds^2 = \alpha^2 \left(- dt^2 + dy^2 \right) + r^2(x) d\Omega^2.
\end{equation}
We can then transform to double null coordinates using
\begin{equation} \label{t y from u v}
t = \frac{v+u}{2},
\qquad
y = \frac{v-u}{2},
\end{equation}
for which
\begin{equation}
ds^2 = - \alpha^2 du dv + r^2 d\Omega^2.
\end{equation}
Comparing to the metric in (\ref{metric}), we have $\sigma = \ln \alpha$ and $r$ is the same. 

The relationship between the radial coordinates $x$ and $y$ is obtained by integrating (\ref{dy dx}),
\begin{equation} \label{y from x}
y = \int dx \, \frac{a(x)}{\alpha(x)}.
\end{equation}
From (\ref{t y from u v}), note that $y_0 \equiv (v_i - u_i)/2$ corresponds to the origin of the computational domain. We can always choose the integration constant in (\ref{y from x}) such that $y_0$ corresponds to a specific value of $x$.

To construct the initial data, we proceed as follows. For any $u,v$ values on an initial hypersurface, we have the corresponding value of $y$ from (\ref{t y from u v}). We then solve for the corresponding value of $x$ from (\ref{y from x}) for a specific static solution. With $x$, we can compute the value of any field for the static solution.

%--------------------------------------------------------------------

\subsection{Procedure}

We use the following procedure to dynamically evolve wormholes:
\begin{itemize}
\item We derive the gravitational field equations and the equations of motion for any matter in the system using the static metric in (\ref{static metric}).
\item We self-consistently solve the gravitational field equations and equations of motion for a static wormhole solution.
\item We transform the static solution to double null coordinates, using the method explained at the end of section \ref{sec:initial data}. We use the transformed solution as our initial data.
\item We place the initial data on the initial hypersurfaces of our computational domain. More specifically, field values from the transformed static solution are used at the grid points on the initial hypersurfaces.
\item To dynamically evolve the system we use the equations of motion, written in double null coordinates, and the evolution equations in (\ref{sigma r evo}). We then self-consistently solve this system of equations using the numerical scheme described in section \ref{sec:numerical methods}. The result is the dynamical evolution of the initial static solution.
\end{itemize}
We have confirmed that the code we have written to perform the dynamical evolution exhibits second order convergence and convergence to a consistent solution. In appendix \ref{app:code tests}, we review how this can be done and present some tests of our code.

%====================================================================

\section{Ellis-Bronnikov wormholes}
\label{sec:EB}

Ellis-Bronnikov wormholes are a family of static spherically symmetric wormhole solutions which can be found analytically \cite{Ellis:1973yv, Bronnikov:1973fh}. They are sourced by a real massless ghost scalar field. As mentioned in the introduction, a  ghost field has the opposite sign in front of the kinetic term in the Lagrangian when compared to a regular field. We review the static wormhole solutions in section \ref{sec:EB static sols}. In section \ref{sec:EB dynamical eqs}, we list the dynamical equations which complement those given in section \ref{sec:double null} before discussing how the static solutions are used as initial data in our simulations in section \ref{sec:EB initial data}. We present results for the expansion and collapse of Ellis-Bronnikov wormholes in section  \ref{sec:EB dynamical evo}.

%--------------------------------------------------------------------

\subsection{Static solutions}
\label{sec:EB static sols}

Analytical static solutions may be found using the metric
\begin{equation} \label{EB static metric}
ds^2 = - \frac{1}{a^2(x)} dt^2 + a^2(x) dx^2 + a^2(x)(x_0^2 + x^2) d\Omega^2,
\end{equation}
which has the form of the general static metric in (\ref{static metric}), where $\alpha(x) = 1/a(x)$,
\begin{equation} \label{EB r eq}
r(x) = a(x)\sqrt{x_0^2 + x^2}
\end{equation}
is the areal radius, and $x_0$ is a constant. Using this metric in the Einstein field equations in (\ref{EFE}), we find the field equations
\begin{eqnarray} \label{static EB EFE}
a'' &= \frac{a^{\prime\,2}}{2a} - a \frac{x_0^2}{2(x_0^2 + x^2)^2}
- a' \frac{2 x}{x_0^2 + x^2} 
+ 4\pi a^3 T\indices{^t_t}
\nonumber \\
a^{\prime\,2} &= - a^2 \frac{x_0^2}{(x_0^2 + x^2)^2}
- 8\pi a^4 T\indices{^x_x}
\nonumber \\
a'' &= 
\frac{a^{\prime \, 2}}{a} 
- \frac{2 x}{x_0^2 + x^2} a'
- 4\pi a^3 \left(-T\indices{^t_t} + T\indices{^x_x} + 2 T\indices{^\theta_\theta} \right),
\end{eqnarray}
where a prime denotes an $x$ derivative.

The real massless ghost scalar field has Lagrangian
\begin{equation} 
\mathcal{L} = - \frac{\eta}{2} (\nabla_\mu \phi)(\nabla^\mu \phi),
\end{equation}
where $\phi$ is the scalar field and $\eta = -1$ for a ghost field. The Lagrangian is minimally coupled to gravity, $\mathcal{L} \rightarrow \sqrt{-\det(g_{\mu\nu})} \, \mathcal{L}$. The equation of motion is
\begin{equation} \label{static EB eom}
\nabla_\mu \nabla^\mu \phi = 0,
\end{equation}
which is the massless Klein-Gordon equation. The energy-momentum tensor has nonvanishing components
\begin{equation} \label{static EB EM}
T\indices{^t_t}
= -T\indices{^x_x}
= T\indices{^\theta_\theta}
= T\indices{^\phi_\phi}
= - \frac{\eta}{2 a^2} \phi^{\prime \, 2}.
\end{equation}

We now proceed to solve the field equations in (\ref{static EB EFE}) and the equation of motion in (\ref{static EB eom}). From (\ref{static EB EM}), we have $-T\indices{^t_t} + T\indices{^x_x} + 2 T\indices{^\theta_\theta} = 0$, which allows the bottom equation in (\ref{static EB EFE}) to be written as
\begin{equation}
\partial_x \left[ (x_0^2 + x^2) \frac{a'}{a} \right] = 0.
\end{equation}
The solution is 
\begin{equation} \label{EB a soln}
a(x) = a_2 \exp \left[ -a_1 \tan^{-1}(x/x_0) \right],
\end{equation}
where $a_1$ and $a_2$ are integration constants. The equation of motion in (\ref{static EB eom}) for the metric in (\ref{EB static metric}) becomes
\begin{equation} \label{static eom}
\partial_x \left[ (x_0^2 + x^2) \phi' \right] = 0.
\end{equation}
The solution is
\begin{equation} \label{EB phi soln}
\phi(x) = \phi_1 \tan^{-1} (x/x_0) + \phi_2,
\end{equation}
where $\phi_1$ and $\phi_2$ are integration constants. There remains the top two equations in (\ref{static EB EFE}). Plugging the results into these equations, we find both are solved for 
\begin{equation}
1 + a_1^2 + 4\pi \eta \phi_1^2 = 0,
\end{equation}
which relates the constants $a_1$ and $\phi_1$. We choose to solve for $\phi_1$,
\begin{equation} \label{phi1 constraint}
\phi_1 = \sqrt{\frac{1 + a_1^2}{4\pi(-\eta)}}.
\end{equation}
As $\phi_1$ must be real, since $\phi$ is a real scalar field, we find that we only have a solution for $\eta = -1$, and hence for a ghost field.

We have solved the field equations and the equation of motion analytically, with the solution given in (\ref{EB a soln}), (\ref{EB phi soln}), and (\ref{phi1 constraint}). To see that this solution describes a wormhole, we can show that the areal radius is nonzero at its minimum. The areal radius is given in (\ref{EB r eq}). Setting the derivative equal to zero, we find that there is an extremum at
\begin{equation}
x_{\mathrm{th}} = a_1 x_0.
\end{equation}
We'll see shortly that this is the location of the wormhole throat. It is straightforward to show that the second derivative evaluated at $x = x_\mathrm{th}$ is positive as long as $a_2$ and $x_0$ have the same sign. We shall assume as much and the extremum is a minimum. The areal radius at its minimum is equal to
\begin{equation}
r_\mathrm{th} = r(x_\mathrm{th}) = a_2 x_0 \sqrt{1 + a_1^2} \, e^{-a_1 \tan^{-1}(a_1)}.
\end{equation}
We find that the areal radius is nonzero and positive at its minimum and we have shown that we have a wormhole. We refer to the minimum value of the areal radius, $r_\mathrm{th}$, as the wormhole throat radius and the wormhole throat is located at $x = x_\mathrm{th}$.

The Misner-Sharp mass function, $m(x)$, gives the total mass inside a radius $r(x)$. It is defined in (\ref{Misner-Sharp def}). Using the metric in (\ref{EB static metric}), the areal radius in (\ref{EB r eq}), and the solution for $a(x)$ in (\ref{EB a soln}), the mass function is
\begin{equation}
m(x) = \frac{a_2 x_0}{2\sqrt{x_0^2 + x^2}} \left[2 a_1 x + x_0(1 - a_1)^2 \right]
\exp \left[ -a_1 \tan^{-1}(x/x_0) \right].
\end{equation}
The ADM mass, $M_\pm$, as viewed from either side of the wormhole, can be found by taking the limit $x\rightarrow \pm \infty$ of $m(x)$,
\begin{equation} \label{EB ADM mass}
M_\pm =
\lim_{x\rightarrow \pm \infty} m(x) = \pm a_1 a_2 x_0 e^{\mp a_1 \pi/2}.
\end{equation}
We will comment on this result shortly.

The Ellis-Bronnikov family of wormhole solutions is given by (\ref{EB a soln}), (\ref{EB phi soln}), and (\ref{phi1 constraint}). It is parameterized in terms of the constants $a_1$, $a_2$, $\phi_1$, and $\phi_2$. We can fix all but one of these constants. A look at the energy-momentum tensor in (\ref{static EB EM}) and the equation of motion in (\ref{static eom}) shows that it is only the derivative $\phi'$ that shows up in the equations. As a consequence, $\phi_2$ does not play a role and we can set $\phi_2 = 0$ without loss of generality. To fix $a_2$, first note that we would like our wormhole solution to be asymptotically flat on both sides. A look at the metric in (\ref{EB static metric}) suggests that this will be the case if $a \rightarrow 1$ in the limits $x \rightarrow \pm \infty$. Taking these limits of (\ref{EB a soln}), we find
\begin{equation}
\lim_{x\rightarrow \pm \infty} a(x) = a_2 e^{\mp a_1\pi/2}.
\end{equation}
Now, the system has a symmetry. Under the scaling 
\begin{equation}
t \rightarrow \sqrt{\lambda} \, t
\qquad
x \rightarrow \frac{x}{\sqrt{\lambda}},
\qquad
x \rightarrow \frac{x_0}{\sqrt{\lambda}},
\qquad
a \rightarrow \lambda a,
\end{equation}
and $\phi$ unchanged, where $\lambda$ is a positive constant, the metric, field equations, and equation of motion are invariant. Using $\lambda = \exp(\pm a_1\pi/2)/a_2$, we can always transform the $\pm$ side such that $a \rightarrow 1$. This shows that both sides of the wormhole are always asymptotically flat. Further, this scaling allows us to transform $a_2$ to any nonzero value. We choose to fix
\begin{equation}
a_2 = 1
\end{equation}
as our convention. With a positive value for $a_2$, we note that $x_0$ is then positive. Finally, $\phi_1$ is determined from $a_1$ from (\ref{phi1 constraint}). The family of solutions is now parameterized in terms of $a_1$ alone.

The Ellis-Bronnikov wormhole solutions are now
\begin{eqnarray} \label{EB static solution}
a(x) &= \exp \left[ -a_1 \tan^{-1}(x/x_0) \right]
\nonumber \\
\phi(x) &= \phi_1 \tan^{-1} (x/x_0)
\nonumber \\
\phi_1 &= \sqrt{\frac{1 + a_1^2}{4\pi(-\eta)}}.
\end{eqnarray}
Consider the wormhole with $a_1 = 0$, for which $a(x) = 1$. Returning to the ADM mass in (\ref{EB ADM mass}), we find that this wormhole is massless. From (\ref{EB r eq}), the areal radius is $r(x) = \sqrt{x_0^2 + x^2}$, which is symmetric about $x_\mathrm{th} = 0$. The $a_1 = 0$ solution is the symmetric massless wormhole. The other solutions are asymmetric and massive.

\begin{figure} 
\centering
\includegraphics[width=6.5in]{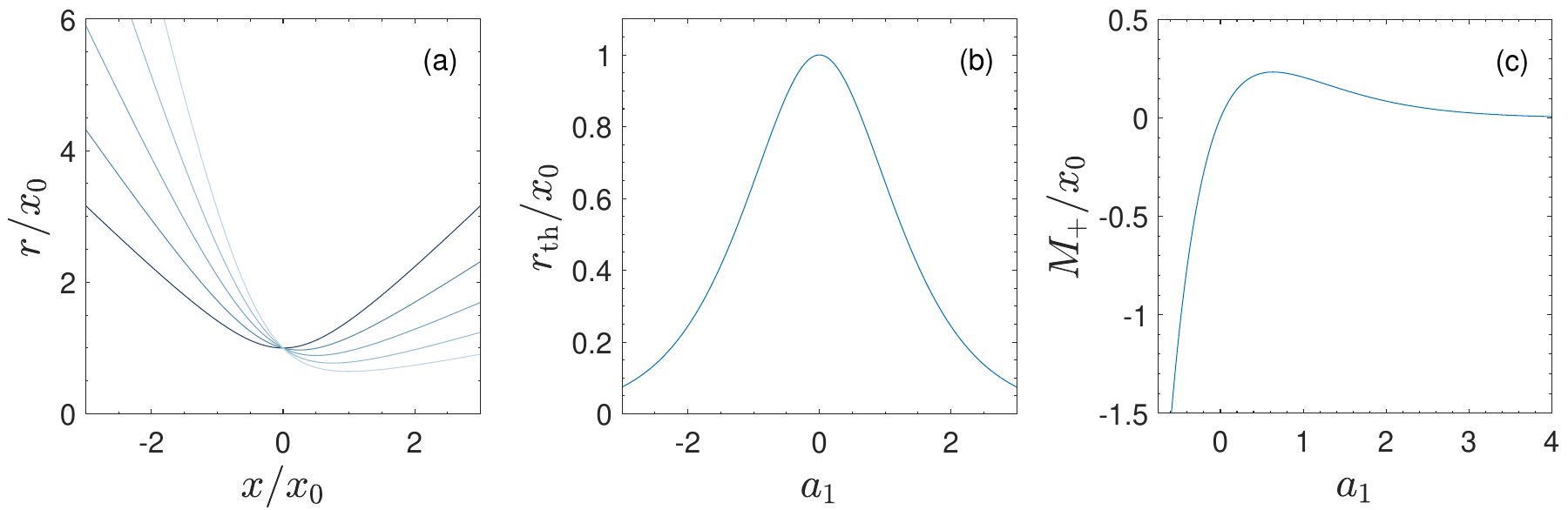} 
\caption{These diagrams display static Ellis-Bronnikov wormhole solutions which are parametrized in terms of the constant $a_1$. (a) The areal radius, $r$, as a function of the radial coordinate, $x$, for (from darkest curve to lightest) $a_1 = 0$, 0.25, 0.5, 0.75, and 1. (b) The wormhole throat radius, $r_\mathrm{th}$, as a function of $a_1$. (c) The ADM mass as viewed from the positive side, $M_+$, as a function of $a_1$.}
\label{fig:static EB}
\end{figure}

In figure \ref{fig:static EB}(a), we show the areal radius, $r(x)$, for $a_1 = 0$, 0.25, 0.5, 0.75, and 1. As $a_1$ increases from zero, the wormhole becomes more asymmetric and the wormhole throat radius decreases. The areal radius for negative $a_1$ can be found from $r(x,a_1) = r(-x,-a_1)$. Figure \ref{fig:static EB}(b) displays the wormhole throat radius, $r_\mathrm{th}$, as a function of $a_1$. We can see that $r_\mathrm{th}$ is largest for $a_1 = 0$. The ADM mass as viewed from the positive side, $M_+$, is displayed in figure \ref{fig:static EB}(c) as a function of $a_1$. $M_+$ is asymptotically zero in the limit $a_1 \rightarrow \infty$, in addition to equaling zero at $a_1 = 0$. $M_-$ can be found from $M_-(a_1) = M_+(-a_1)$. $M_\pm$ have maximum values at $a_1 = \pm 2/\pi$. In both cases, the maximum mass equals $2/\pi e$.

%--------------------------------------------------------------------

\subsection{Dynamical equations}
\label{sec:EB dynamical eqs}

To dynamically evolve the Ellis-Bronnikov wormhole, we use double null coordinates, as we discussed in section \ref{sec:double null}. We may therefore use the field equations in (\ref{constraint eqs}) and (\ref{sigma r evo}). Additionally, we need equations describing the matter sector.

For the Ellis-Bronnikov wormhole, we have a real massless ghost scalar field, $\phi$, which sources the wormhole geometry. We will also include a regular real massless scalar field, $\chi$, which we will use to perturb the system. The Lagrangian describing the matter sector is then
\begin{equation} 
\mathcal{L} = 
- \frac{\eta}{2} (\nabla_\mu \phi)(\nabla^\mu \phi)
- \frac{1}{2} (\nabla_\mu \chi)(\nabla^\mu \chi),
\end{equation}
where $\eta = -1$ so that $\phi$ is a ghost field. We minimally couple this to gravity, $\mathcal{L} \rightarrow \sqrt{-\det(g_{\mu\nu})} \, \mathcal{L}$. The equations of motion are two copies of the massless Klein-Gordon equation, 
\begin{equation} 
\nabla_\mu \nabla^\mu \phi = 0,
\qquad
\nabla_\mu \nabla^\mu \chi = 0.
\end{equation}
For the double null metric in (\ref{metric}), the equations of motion work out to
\begin{equation} \label{phi chi eom}
\phi_{,uv}
= - \frac{1}{r} \left(r_{,u} \phi_{,v} +  r_{,v} \phi_{,u} \right),
\qquad
\chi_{,uv}
= - \frac{1}{r} \left(r_{,u} \chi_{,v} +  r_{,v} \chi_{,u} \right).
\end{equation}
The energy-momentum tensor is given by
\begin{equation}
T_{\mu\nu} = T^\phi_{\mu\nu} + T^\chi_{\mu\nu},
\end{equation}
where $T^\phi_{\mu\nu}$ is the contribution from the ghost field and $T^\chi_{\mu\nu}$ is the contribution from the regular field. The components are
\begin{eqnarray} \label{EB T phi}
T_{uu}^\phi &= \eta \, \phi_{,u}^2
\nonumber \\
T_{vv}^\phi &= \eta \, \phi_{,v}^2
\nonumber \\
T_{uv}^\phi &= 0
\nonumber \\
T_{\theta\theta}^\phi &= \eta \, 2 r^2 e^{-2\sigma} \phi_{,u} \phi_{,v}
\end{eqnarray}
and
\begin{eqnarray} \label{EB T chi}
T_{uu}^\chi &= \chi_{,u}^2
\nonumber \\
T_{vv}^\chi &= \chi_{,v}^2
\nonumber \\
T_{uv}^\chi &= 0
\nonumber \\
T_{\theta\theta}^\chi &= 2 r^2 e^{-2\sigma} \chi_{,u} \chi_{,v},
\end{eqnarray}
along with $T_{\phi\phi}^\phi = T_{\theta\theta}^\phi \sin^2\theta$ and $T_{\phi\phi}^\chi = T_{\theta\theta}^\chi \sin^2\theta$, with the rest of the components vanishing due to spherical symmetry.

%--------------------------------------------------------------------

\subsection{Initial data}
\label{sec:EB initial data}

To construct initial data from a static wormhole solution, we follow the methods described in section \ref{sec:initial data}. We note that the computation of the new radial coordinate in (\ref{y from x}) becomes
\begin{equation} \label{EB y x int}
y = \int dx \, a^2(x) = \int dx \, \exp \left[ - 2a_1 \tan^{-1}(x/x_0) \right]
\end{equation}
and that we choose the integration constant such that $y = y_0 = (v_i - u_i)/2$ when $x = x_\mathrm{th}$. This places the wormhole throat at the origin of the computational domain.

The initial data must be computed numerically, because of the integral in (\ref{EB y x int}), except for the case of the symmetric massless wormhole with $a_1 = 0$. For the symmetric massless wormhole, the initial data can be written down analytically and is
\begin{eqnarray}
\sigma(u,v) &= 0
\nonumber \\
r(u,v) &= \frac{1}{2} \sqrt{4x_0^2 + (v-u)^2}
\nonumber \\
\phi(u,v) &= \frac{1}{\sqrt{4\pi}} \tan^{-1} \left( \frac{v - u}{2x_0} \right).
\end{eqnarray}

\begin{figure} 
\centering
\includegraphics[width=6.5in]{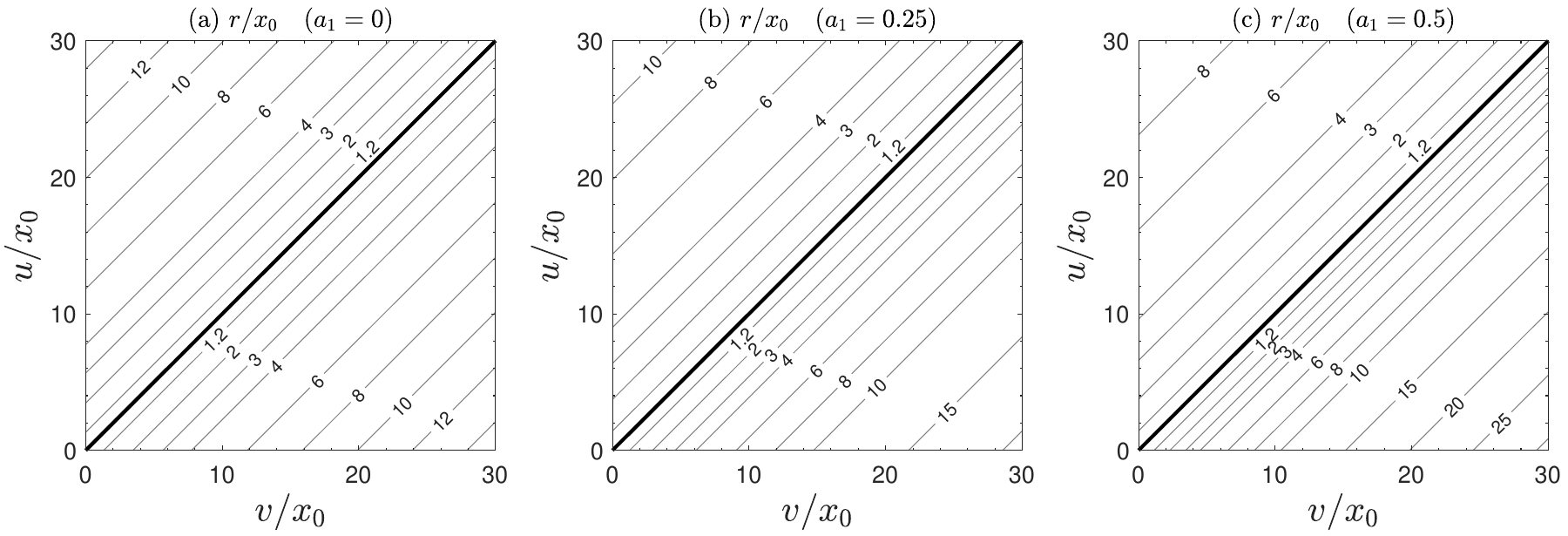} 
\caption{Contour diagrams for the areal radius, $r$, in double null coordinates for static Ellis-Bronnikov wormholes defined by (a) $a_1 = 0$, (b) 0.25, and (c) 0.5. In each diagram, the gray lines are contours with values of $r$ as labeled and the thick black lines mark the wormhole throat with radii (a) $r_\mathrm{th}/x_0 = 1$, (b) 0.970, and (c) 0.887.}
\label{fig:static EB uv}
\end{figure}

When dynamically evolving a system using double null coordinates, one of the most useful diagrams for understanding how the system evolves is the contour diagram for the areal radius. To get a sense for this diagram, we display it for various static solutions in figure \ref{fig:static EB uv}. Figure \ref{fig:static EB uv} is therefore analogous to figure \ref{fig:static EB}(a), but in double null coordinates. Figure \ref{fig:static EB uv}(a) is for the symmetric massless wormhole with $a_1 = 0$. The thick black line is the contour for $r = r_\mathrm{th} = x_0$. The two sides of the thick black line therefore plot the two sides of the wormhole. The gray lines are contours for various values of $r$ as labeled. Note that they are symmetric about the wormhole throat. Figures \ref{fig:static EB uv}(b) and \ref{fig:static EB uv}(c) are for asymmetric wormholes with $a_1 = 0.25$ and $0.5$;~the thick black lines are contours for $r = r_\mathrm{th} = 0.970 x_0$ and $0.887 x_0$, respectively. Looking at the gray lines, we can get a sense of the asymmetry of these wormholes.

It follows from (\ref{t y from u v}) that lines of constant $v - u$ are lines of constant $y$ and hence lines of constant $x$. For this reason, static systems have straight contour lines with unit slope, as can be seen in figure \ref{fig:static EB uv}. As a consequence, if we were to perform a dynamical evolution using the static solution as initial data and the dynamical evolution were to reproduce the straight contours in figure \ref{fig:static EB uv}, we would conclude that the static solution is stable with respect to time dependent perturbations since the system would be maintaining the static configuration. As we'll see, the dynamically evolved system will initially maintain the static configuration, before evolving away from it.

For the dynamical evolution, the initial data is placed on the $u = u_i$ and $v = v_i$ hypersurfaces, which are the axes in figure \ref{fig:static EB uv}. We can use a static solution as initial data or we can  augment the static solution with a pulse of regular scalar field. The pulse acts as an explicit perturbation. Physically, we can imagine sending a pulse of scalar field into the wormhole. Pulses may be placed at a single or multiple locations on both initial hypersurfaces. To keep things simple, we consider a single pulse placed on the $u = u_i$ hypersurface. Since the wormhole throat is placed at the origin of the computational domain, this corresponds to placing the pulse at $x > x_\mathrm{th}$.

In the dynamical equations, only derivatives of $\chi$ show up. We therefore choose to define the pulse in terms of its derivative, 
\begin{equation} \label{chi pulse 1}
\chi_{,v} (u_i, v) =  \frac{A}{x_0} \sin^2 \left( \pi \frac{v - v_1}{v_2-v_1} \right),
\end{equation}
for $v_1 < v < v_2$ and zero everywhere else on the initial hypersurfaces, where $A$, $v_1$, and $v_2$ are constants. This is a commonly used form for a pulse because $\chi_{,v}$ and $\chi_{,vv}$ are zero at $v = v_1$ and $v = v_2$. Although only derivatives of $\chi$ show up in the dynamical equations, the numerical scheme we use to solve the equations makes use of $\chi$, which is
\begin{equation} \label{chi pulse 2}
\chi(u_i, v)
= \left\{
\begin{array}{ll}
0 & \textrm{for } v < v_1
\\ [10pt]
\displaystyle
\frac{A}{4\pi x_0} \left[
2\pi (v- v_1) - (v_2 - v_1) \sin\left( 2\pi \frac{v - v_1}{v_2 - v_1} \right) 
\right]
& \textrm{for } v_1 < v < v_2
\\[15pt]
\displaystyle
\frac{A}{4\pi x_0} \left[2\pi (v_2- v_1) \right]
& \textrm{for } v > v_2.
\end{array} \right.
\end{equation}
The factor of $x_0$ is included for convenience, since it can be used to form the dimensionless variable $v/x_0$.

If we include the pulse, the initial data on the $v = v_i$ hypersurface is unchanged from the static solution, but the initial data on the $u = u_i$ hypersurface is modified since the energy-momentum tensor now includes a contribution from $T^\chi_{\mu\nu}$. More specifically, $\sigma$ and $r$ may change and must be computed, while $\phi$ is unchanged. There are two standard options. In the first option, we keep $\sigma(u_i,v)$ unchanged from the static solution and compute $r(u_i,v)$ using the second constraint equation in (\ref{constraint eqs}). In the second option, we instead keep $r(u_i,v)$ unchanged and compute $\sigma(u_i,v)$ using the same constraint equation. Both options work well and both are allowed because of the coordinate gauge freedom in double null coordinates. We choose the first option. 

To solve the constraint equation on the $u = u_i$ initial hypersurface, we need $\sigma_{,v}(u_i,v)$ and $\phi_{,v}(u_i,v)$, which are unchanged from the static solution, and $r(u_i,v_i)$ and $r_{,v}(u_i,v_i)$ at the origin, which are also unchanged from the static solution since we will only consider pulses outside the origin. The derivatives can be computed analytically,
\begin{equation} \label{sigmaV rV phiV}
\sigma_{,v} = \frac{a_1 x_0}{2 a^2(x)(x_0^2 + x^2)},
\quad
r_{,v} = \frac{x - a_1 x_0}{2 a(x) \sqrt{x_0^2 + x^2}},
\quad
\phi_{,v} = \frac{\phi_1 x_0}{2a^2(x)(x_0^2 + x^2)}.
\end{equation}
For the symmetric massless case, the derivatives can be transformed to double null coordinates analytically,
\begin{equation}
\sigma_{,v} = 0,
\qquad
r_{,v} = \frac{v - u}{2\sqrt{4x_0^2 + (v - u)^2}},
\qquad
\phi_{,v} = \frac{1}{\sqrt{4\pi}} \frac{2 x_0}{4x_0^2 + (v-u)^2}.
\end{equation}
We can now integrate the constraint equation outward from $v = v_i$ to obtain the updated $r(u_i,v)$.  

%--------------------------------------------------------------------

\subsection{Dynamical evolution}
\label{sec:EB dynamical evo}

We now present results for the dynamical evolution of Ellis-Bronnikov wormholes \cite{Shinkai:2002gv, Gonzalez:2008xk, Doroshkevich:2008xm, Shinkai:2015aqa, Shinkai:2017xkx, Novikov:2019rus, Novikov:2020xwr, Calhoun:2022xrw, Xu:2025jad}. We'll find that the static solutions, when dynamically evolved, expand or collapse. We focus on examples of expansion first and then look at examples of collapse.

The static solutions we use for initial data depend on the parameters $x_0$ and $a_1$. We can absorb $x_0$ into a redefinition of variables so that it does not have to be specified. Throughout this subsection, we use the scaled variables
\begin{eqnarray}
u \rightarrow \frac{u}{x_0},
\qquad
v \rightarrow \frac{v}{x_0},
\qquad
r \rightarrow \frac{r}{x_0},
%\qquad
%m \rightarrow \frac{m}{x_0},
\qquad
T_{uu} \rightarrow x_0^2 T_{uu},
\qquad
T_{vv} \rightarrow x_0^2 T_{vv}.
\end{eqnarray}

%++++++++++++++++++++++++++++++++++++++++++++++++++++++

\subsubsection{Wormhole expansion}

\begin{figure} 
\centering
\includegraphics[width=6.5in]{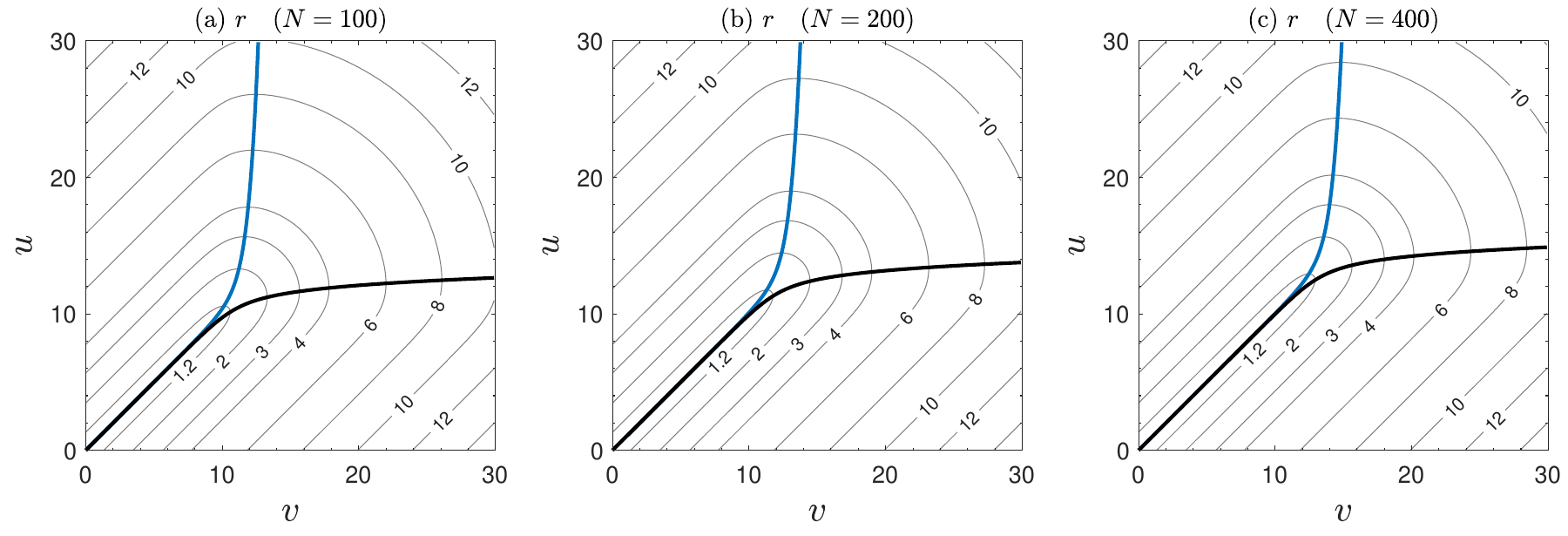} 
\caption{Contour diagrams for the areal radius, $r$, for the dynamical evolution of the symmetric massless Ellis-Bronnikov wormhole. Each diagram displays an expanding wormhole. The black curves display the apparent horizon $r_{,u} = 0$ and the blue curves display the apparent horizon $r_{,v} = 0$. Each diagram uses the same initial data, but evolves the system with grid spacing $\Delta u = \Delta v = 1/N$ with (a) $N = 100$, (b) 200, and (c) 400. As the grid spacing decreases, the evolution maintains the static configuration longer. This is the expected behavior for a unstable static solution.}
\label{fig:EB r symmetric no pulse}
\end{figure}

We begin with the simplest possibility:~the evolution of the symmetric massless wormhole without a scalar field pulse. The contour diagram for the areal radius is shown in figure \ref{fig:EB r symmetric no pulse}(a). The static solution is placed on the initial hypersurfaces, which are the axes in the figure, and the bulk of the diagram displays the evolution. The gray lines are contours for $r$, with the values for $r$ as labeled, and the thick black and blue lines are apparent horizons defined by $r_{,u} = 0$ and $r_{,v} = 0$, respectively.

One of the first things to notice is that the diagram looks different than the diagram for the static solution in figure \ref{fig:static EB uv}(a), indicating that the system evolves away from the static solution. To understand the evolution, it helps to begin in the bottom left corner. Initially, as the system evolves away from this corner, figure \ref{fig:EB r symmetric no pulse}(a) is identical to figure \ref{fig:static EB uv}(a), indicating that the system is maintaining the static configuration. The apparent horizons are on top of one another and mark the wormhole throat, since for the static solution the wormhole throat is defined by $r_{,x} = 0$, which transforms to $r_{,y} = r_{,u} = r_{,v} = 0$. At around $u \approx 10$, the apparent horizons separate and the system evolves away from the static configuration. In between the apparent horizons, moving in the timelike direction necessarily moves across increasing radii. This signifies that the wormhole is expanding since the minimum radius is increasing. As we approach the top right corner, the wormhole appears to expand indefinitely. The apparent horizons evolve to being asymptotically vertical and horizontal, indicating the presence of cosmological horizons at roughly $u/x_0 \approx 13$ and $v/x_0 \approx 13$.

The evolution in figure \ref{fig:EB r symmetric no pulse}(a) does not contain a scalar field pulse. The only perturbation present is the discretization error inherent to the code. The discretization error can be reduced, which decreases the strength of the perturbation, by including more grid points so that the discrete grid better approximates the continuum. Figure \ref{fig:EB r symmetric no pulse}(a) is made with $N = 100$ grid points per unit interval, with the spacing between grid points equal to $\Delta u = \Delta v = 1/N$. Figures \ref{fig:EB r symmetric no pulse}(b) and \ref{fig:EB r symmetric no pulse}(c) use the same initial data, but evolve the system using $N = 200$ and $400$ grid points per unit interval. We find that as the strength of the perturbation is reduced, the system maintains the static configuration longer before eventually evolving away. This is the expected behavior when the static solution is unstable with respect to time dependent perturbations \cite{Shinkai:2002gv}. Indeed, static Ellis-Bronnikov wormholes have been shown to be unstable using a linear stability analysis \cite{Gonzalez:2008wd}.

One can compute robust examples of wormhole expansion---examples which change negligibly when the grid spacing is reduced---by explicitly perturbing the system with a pulse of ghost scalar field (instead of a pulse of regular scalar field) \cite{Shinkai:2002gv, Gonzalez:2008xk, Doroshkevich:2008xm, Calhoun:2022xrw, Xu:2025jad}. We do not consider such perturbations here because the examples in figure \ref{fig:EB r symmetric no pulse} exhibit the properties of wormhole expansion well enough for our purposes. Further, we will not have the option to perturb with a pulse of ghost field when we study the quantum corrected Schwarzschild black hole in the next section, making figure \ref{fig:EB r symmetric no pulse} the relevant diagram for comparison.

\begin{figure} 
\centering
\includegraphics[width=6.5in]{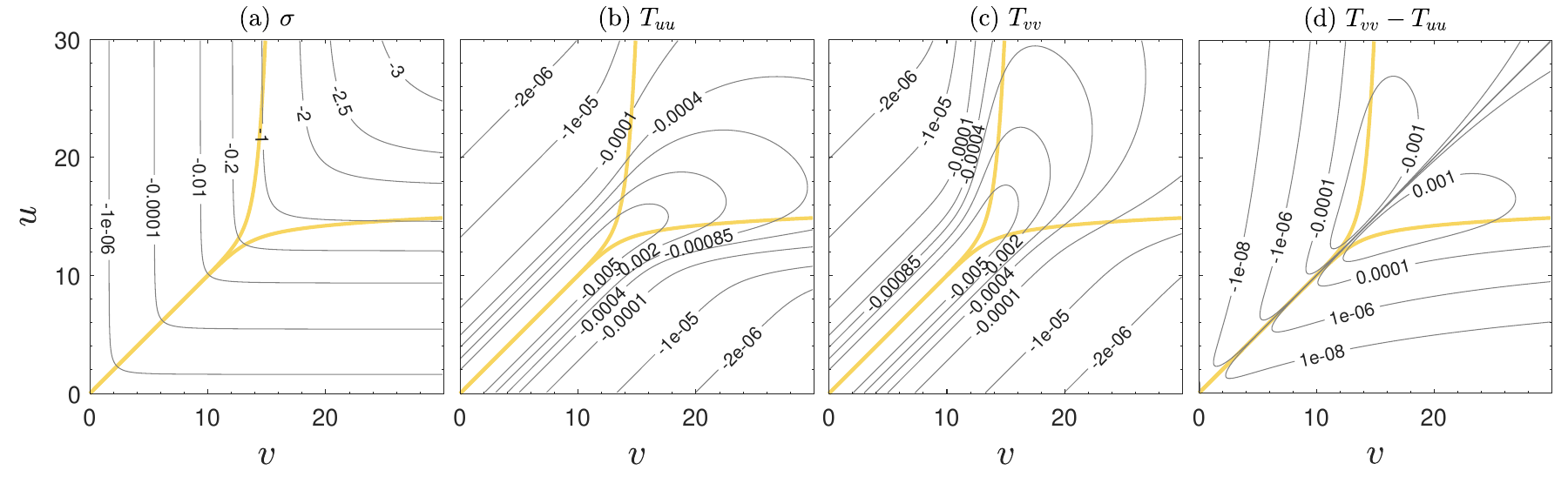} 
\caption{Each diagram is for the same dynamical evolution shown in figure \ref{fig:EB r symmetric no pulse}(c). The yellow curves are the apparent horizons from figure \ref{fig:EB r symmetric no pulse}(c). The diagrams display contours for (a) the metric field $\sigma$, (b) the outgoing energy-momentum $T_{uu}$, (c) the ingoing energy-momentum $T_{vv}$, and (d) the net flow of energy momentum, $T_{vv} - T_{uu}$.}
\label{fig:EB sigma T no pulse}
\end{figure}

The metric in double null coordinates is parameterized by $r(u,v)$ and $\sigma(u,v)$. We show the contour diagram for $\sigma$ in figure \ref{fig:EB sigma T no pulse}(a) for the same evolution shown in figure \ref{fig:EB r symmetric no pulse}(c). The apparent horizons are included in yellow for convenience. Along the axes $\sigma = 0$, which is the value for the static solution. Note that near the axes the contours are not straight lines with unit slope. We explained in section \ref{sec:EB initial data} that we expect such contours when the dynamical evolution is maintaining the static configuration. We do not see them in figure \ref{fig:EB sigma T no pulse}(a) because $\sigma = 0$ in the static solution. We will show an example below of the dynamical evolution of an asymmetric static solution which has nonzero $\sigma$ and we will find straight contours with unit slope near the axes, as expected.

Since apparent horizons are defined by $r_{,u} = 0$ and $r_{,v} = 0$, the mass contained inside an apparent horizon of radius $r$ is given by $r/2$, as follows from (\ref{mass function}). Returning to figure \ref{fig:EB r symmetric no pulse}, we can see that once the system moves away from the static configuration, the apparent horizons evolve along increasing radii. We therefore find that for an expanding wormhole, the mass inside the apparent horizons is increasing. For this to happen, we would expect energy to flow inward. We can see if this occurs by looking at the energy-momentum tensor.

Figures \ref{fig:EB sigma T no pulse}(b) and \ref{fig:EB sigma T no pulse}(c) display the energy-momentum tensor components $T_{uu}$ and $T_{vv}$. $T_{uu}$ describes outgoing energy-momentum and $T_{vv}$ describes ingoing energy-momentum. More specifically, focusing on the region $v > u$, a positive value for $T_{uu}$ indicates energy-momentum that is flowing radially outward and a positive value of $T_{vv}$ indicates energy-momentum that is flowing radially inward. A convenient place to look at these values is at future null infinity. This region is located at $v \rightarrow \infty$ and below the cosmological horizon. Near this region, both $T_{uu}$ and $T_{vv}$ are negative. The net flow is given by $T_{vv} - T_{uu}$, which is shown in figure \ref{fig:EB sigma T no pulse}(d). Near future null infinity we find that $T_{vv} - T_{uu}$ is positive and the net flow is ingoing. This is precisely what we would expect to find when the mass inside the apparent horizons is increasing.

\begin{figure} 
\centering
\includegraphics[width=4.5in]{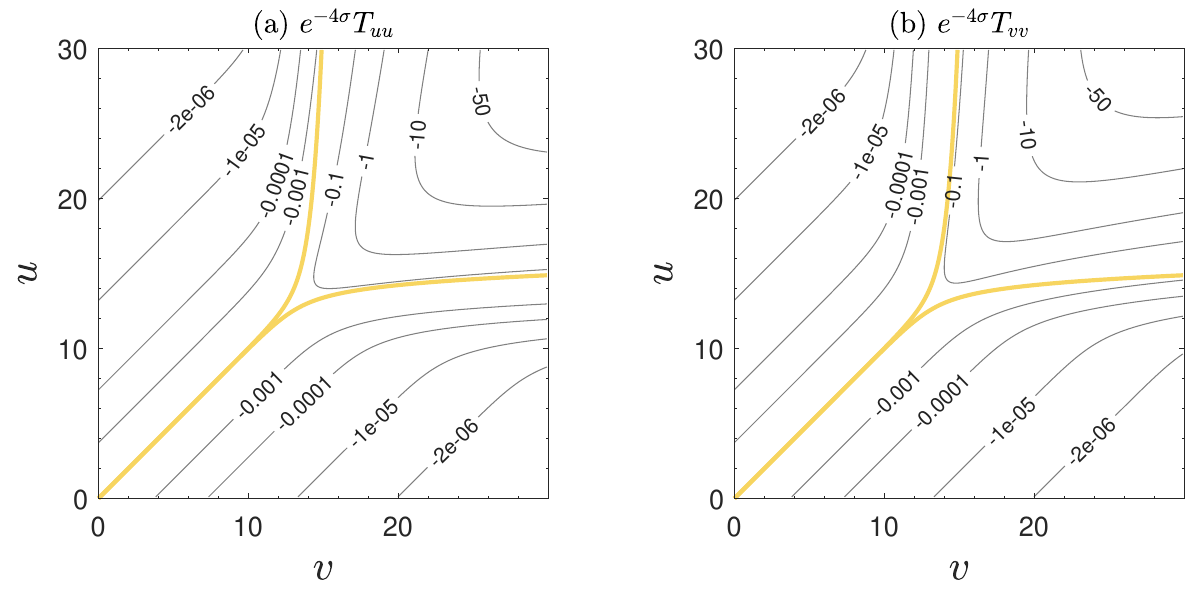} 
\caption{Contour diagrams for the null energy along null geodesics for the same evolution shown in figure \ref{fig:EB r symmetric no pulse}(c) with apparent horizons in yellow.}
\label{fig:EB null energy no pulse}
\end{figure}

As the system evolves, the wormhole expands. Since the wormhole is getting bigger, intuitively we would expect the violation of the null energy condition to worsen and for the null energy to become more negative. To get a sense for this, consider a path through the middle of the separated horizons in figures \ref{fig:EB sigma T no pulse}(b) and \ref{fig:EB sigma T no pulse}(c) for $T_{uu}$ and $T_{vv}$ (since the apparent horizons are symmetric this path would be the line $u = v$). We find that this path moves along values of $T_{uu}$ and $T_{vv}$ which increase, going from initially negative values toward zero. We must remember that $T_{uu}$ and $T_{vv}$ do not give the null energy along null geodesics, as explained in section \ref{sec:Raychaudhuri}. The null energy along null geodesics is given by $e^{-4\sigma}T_{uu}$ and $e^{-4\sigma} T_{vv}$, which we plot in figure \ref{fig:EB null energy no pulse}. Here we find, as expected, along the path through the middle of the separated horizons the null energy decreases, becoming more negative, as the wormhole expands.

\begin{figure} 
\centering
\includegraphics[width=6.5in]{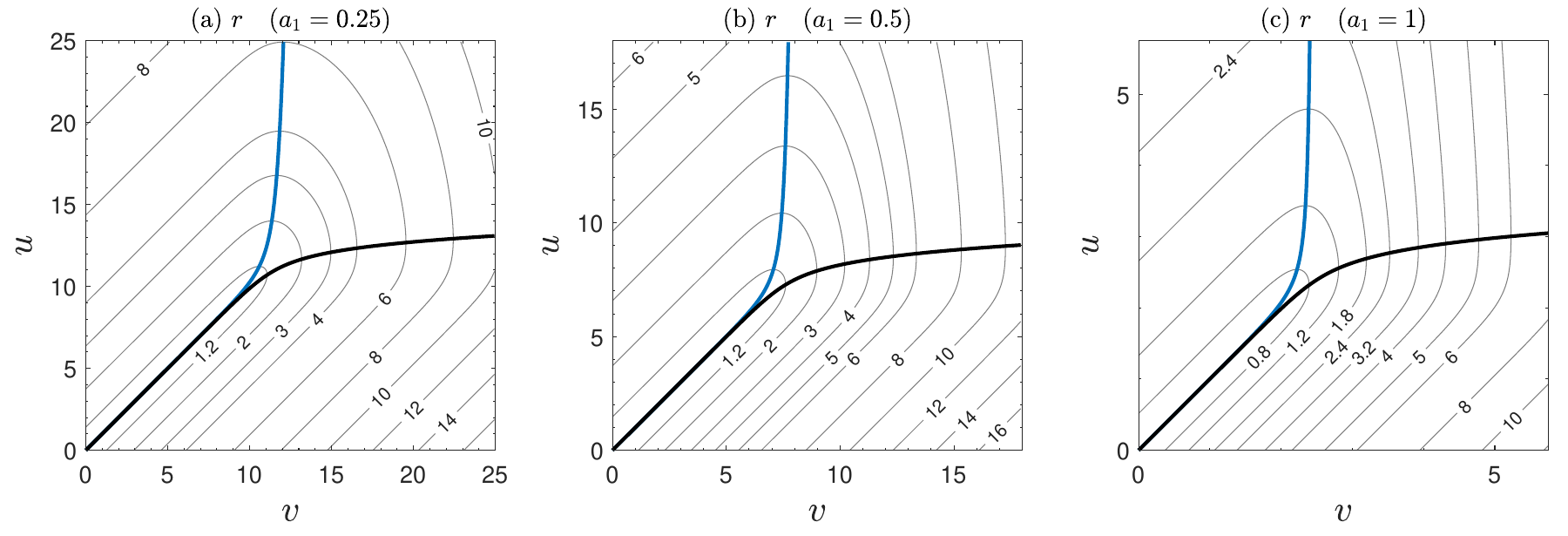} 
\caption{Contour diagrams for the areal radius, $r$, for the dynamical evolution of asymmetric massive Ellis-Bronnikov wormholes defined by (a) $a_1 = 0.25$, (b) 0.5, and (c) 1. The black curves display the apparent horizon $r_{,u} = 0$ and the blue curves display the apparent horizon $r_{,v} = 0$. All diagrams are made using grid spacing $\Delta u = \Delta v = 1/400$.}
\label{fig:EB r asymmetric no pulse}
\end{figure}

So far we have considered the dynamical evolution of the symmetric massless wormhole. We now present examples of asymmetric massive wormholes, but still without a scalar field pulse. We show contour diagrams for the areal radius for a few examples in figure \ref{fig:EB r asymmetric no pulse}. Figures \ref{fig:EB r asymmetric no pulse}(a), \ref{fig:EB r asymmetric no pulse}(b), and \ref{fig:EB r asymmetric no pulse}(c) are for different static solutions, where each static solution is incrementally more asymmetric, since $a_1$ increases. We can see that the resulting simulations, which display wormhole expansion, are also incrementally more asymmetric. Just as in the symmetric massless case, if we decrease the strength of the perturbation by decreasing the spacing between grid points, we have found that the static configurations are maintained longer during the evolution, indicating that the static solutions are unstable.

As the asymmetry increases, we can see from figure \ref{fig:EB r asymmetric no pulse} that the system evolves away from the static configuration sooner. It follows from the energy-momentum tensor in (\ref{EB T phi}) and the equation for $\phi_{,v}$ in (\ref{sigmaV rV phiV}), along with the corresponding equation for $\phi_{,u}$, that null energy along null geodesics for static solutions is equal to $e^{-4\sigma}T_{uu} = e^{-4\sigma} T_{vv} = \eta \phi_1^2x_0^2/4(x_0^2 + x^2)^2$. As we increase $a_1$, we find from (\ref{EB static solution}) that the amplitude for the ghost field, $\phi_1$, increases. As we increase $a_1$ the system therefore begins with a larger violation of the null energy condition. With a larger violation, we are not surprised to find that the system evolves away from the static configuration sooner.

\begin{figure} 
\centering
\includegraphics[width=6.5in]{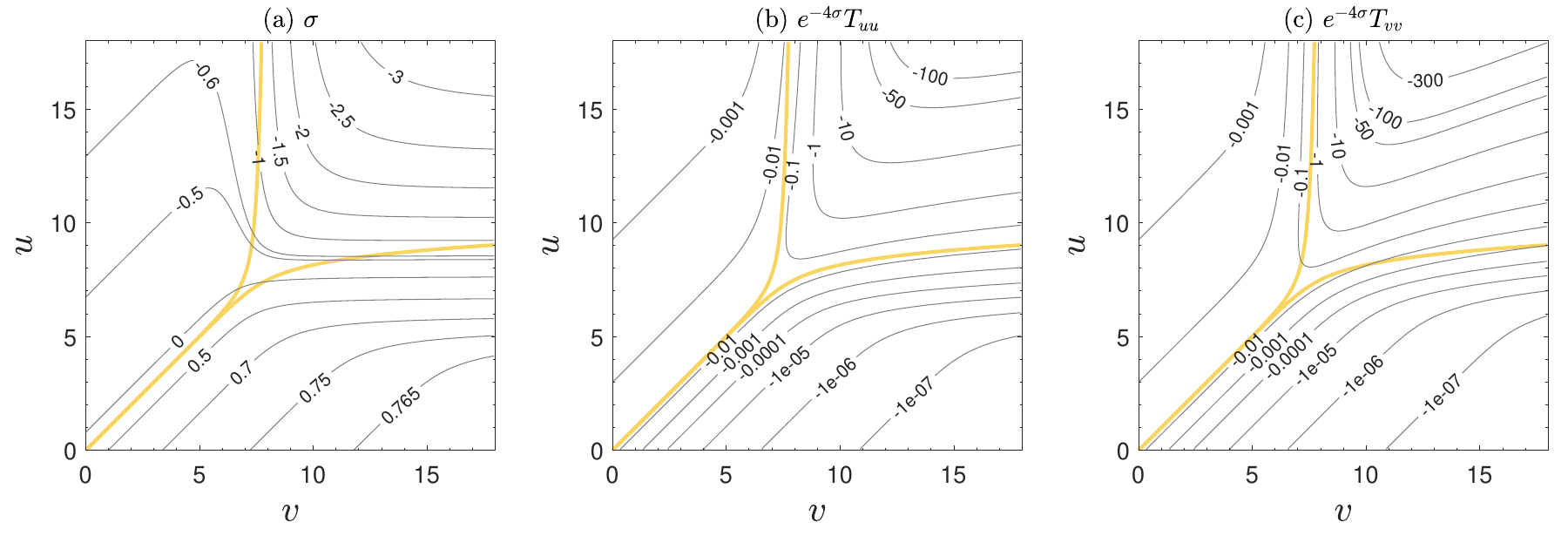} 
\caption{Each diagram is for the same dynamical evolution shown in figure \ref{fig:EB r asymmetric no pulse}(b), with the apparent horizons in yellow. The diagrams display contours for (a) the metric field $\sigma$ and (b, c) the null energy along null geodesics.}
\label{fig:EB sigma T asymmetric no pulse}
\end{figure}

We have studied other properties of the dynamical evolution of the asymmetric massive static solutions. Similarly to the symmetric massless case, we have found that the mass inside the apparent horizons increases and that it can be understood from the components of the energy-momentum tensor. Since we found these properties to be qualitatively similar to the symmetric massless case, we forgo presenting details. One quantity that is qualitatively different is $\sigma$. We show the contour diagram for $\sigma$ in figure \ref{fig:EB sigma T asymmetric no pulse}(a) for the same evolution shown in figure \ref{fig:EB r asymmetric no pulse}(b). Since the asymmetric static solutions have nonzero $\sigma$, we can see in figure \ref{fig:EB sigma T asymmetric no pulse}(a) that the contours are straight lines with unit slope near the axes, as expected. This is in contrast to the symmetric massless solution shown in figure \ref{fig:EB sigma T no pulse}(a). We show in figures \ref{fig:EB sigma T asymmetric no pulse}(b) and \ref{fig:EB sigma T asymmetric no pulse}(c) the null energy along null geodesics, since it depends on $\sigma$. Along a path through the middle of the separated apparent horizons, we can see that the null energy decreases, as expected.

%++++++++++++++++++++++++++++++++++++++++++++++++++++++

\subsubsection{Wormhole collapse}

We now introduce an explicit perturbation in the form of a pulse of regular scalar field. We will find that such a pulse causes the wormhole to collapse. We begin with the symmetric massless static wormhole augmented by a pulse with parameters $A = 0.002$, $v_1 = 1$, and $v_2 = 3$. The pulse is placed on the $u = u_i$ initial hypersurface and introduces an asymmetry into the system. However, these pulse parameters give a relatively weak pulse and the asymmetry will not always be discernible.

\begin{figure} 
\centering
\includegraphics[width=4.5in]{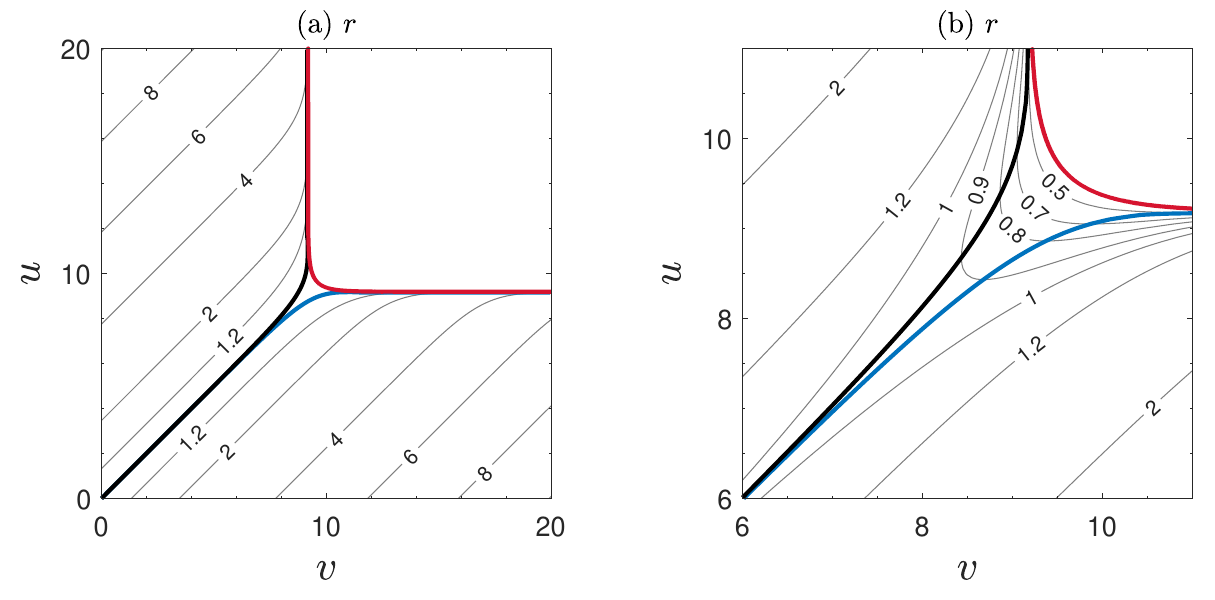} 
\caption{(a) Contour diagram for the areal radius, $r$, for the dynamical evolution of the symmetric massless Ellis-Bronnikov wormhole augmented by a scalar field pulse defined by $A = 0.002$, $v_1 = 1$, and $v_2 = 3$. The black curve displays the apparent horizon $r_{,u} = 0$ and the blue curve displays the apparent horizon $r_{,v} = 0$. The red curve displays the central singularity. The region beyond the red curve is not part of the spacetime. (b) is a magnification of the region between the apparent horizons shown in (a). These diagrams display a collapsing wormhole and the formation of black holes and are made using grid spacing $\Delta u = \Delta v = 1/400$.}
\label{fig:EB r symmetric with pulse}
\end{figure}

We show the contour diagram for the areal radius in figure \ref{fig:EB r symmetric with pulse}. 
Comparing figure \ref{fig:EB r symmetric with pulse}(a) to the static solution in figure \ref{fig:static EB uv}(a), we find that they are different, indicating the system evolves away from the static configuration. We also cannot see any discernible asymmetry introduced by the pulse. To understand the evolution we again start in the bottom left corner and move outward. Initially, figures \ref{fig:EB r symmetric with pulse}(a) and \ref{fig:static EB uv}(a) look the same and the system is maintaining the static configuration. At around $u \approx 7$, the apparent horizons begin to separate. Figure \ref{fig:EB r symmetric with pulse}(b) magnifies the region between the apparent horizons, where we can see, when moving in timelike directions, that the areal radius decreases. Since the minimum radius is decreasing, the wormhole is collapsing. Eventually the radius collapses completely and a singularity forms, which is marked by the red curve. The region beyond the red curve is not part of the spacetime. The apparent horizons, as well as the $r$ contours, evolve to be asymptotically vertical and horizontal, indicating event horizons at $u \approx 9.5$ and $v \approx 9.5$. This signifies the formation of black holes which enclose the mouths of the wormhole.

There are a couple of interesting things we can note. First, the radius of an apparent horizon typically cannot decrease. It can decrease here because the null energy condition is violated \cite{Gonzalez:2008xk}. Second, from figure \ref{fig:EB r symmetric with pulse}(a) we can see that it is possible to travel through the wormhole and return before it collapses. Consider a null ray starting at $(v,u) = (4,0)$ and traveling up. After it crosses the wormhole throat, and hence travels through the wormhole, it can start traveling to the right at, say, $(v,u) = (4,6)$. It will then cross back to the original side of the wormhole, never having crossed the event horizon at $u \approx 9.5$. This collapsing wormhole is therefore temporarily traversable.

\begin{figure} 
\centering
\includegraphics[width=4.5in]{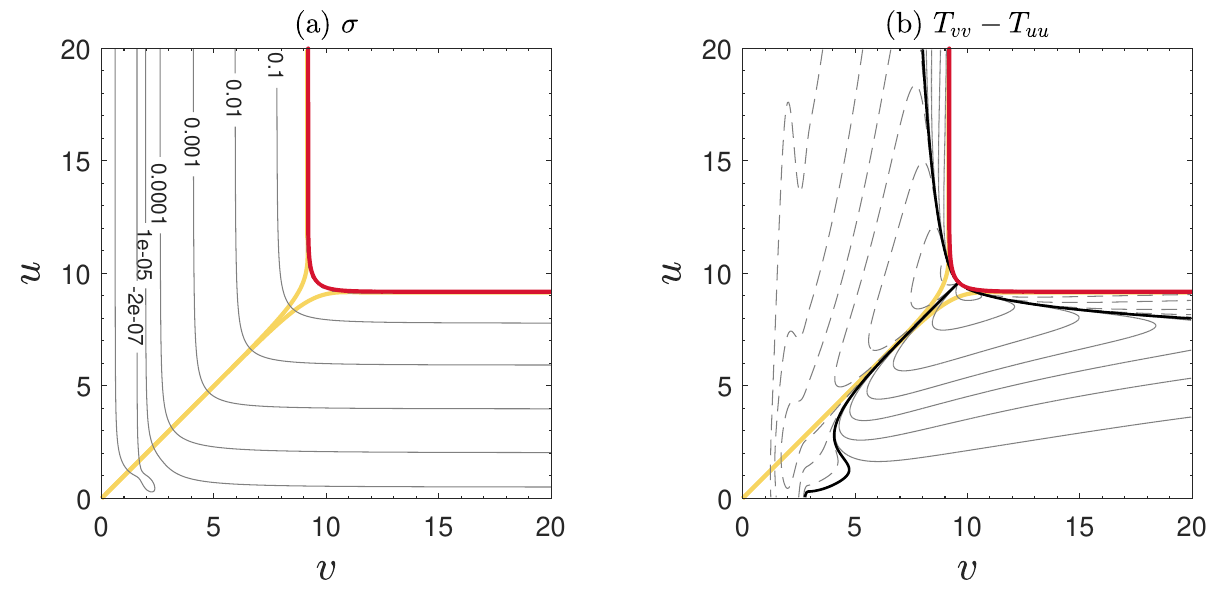} 
\caption{Contours diagrams for (a) the metric field $\sigma$ and (b) the net flow of energy-momentum, $T_{vv} - T_{uu}$, for the same evolution shown in figure \ref{fig:EB r symmetric with pulse}, with the apparent horizons shown in yellow. In (b), solid gray lines have negative contour values, with contours shown in the range $-10^{-7}$ to $-10^{-2}$, and dashed gray lines have positive contour values, with contours shown in the range $+10^{-7}$ to $+10^{-2}$. The thick black lines mark where $T_{vv} - T_{uu} = 0$.}
\label{fig:EB sigma T m with pulse}
\end{figure}

The contour diagram for the metric function $\sigma$ is shown in figure \ref{fig:EB sigma T m with pulse}(a). On the initial hypersurfaces $\sigma = 0$ and we find that $\sigma$ increases during the evolution. This is opposite to the expanding case, where we recall that $\sigma$ decreases, as can be seen in figure \ref{fig:EB sigma T no pulse}(a). Note that near the axes, the contours are not straight lines with unit slope, which would be our expectation when the dynamical evolution is maintaining the static configuration. Just as we explained for the expanding case and figure \ref{fig:EB sigma T no pulse}(a), this occurs because $\sigma = 0$ for the static solution.

Returning to figure \ref{fig:EB r symmetric with pulse}(b), we can see that as the wormhole collapses, the apparent horizons evolve along decreasing radii. As a consequence, the mass inside the apparent horizons decreases. For this to happen, we would expect energy to flow outward. We again find opposite behavior when compared to the expanding case, where we recall that the mass inside the apparent horizons increases. However, the situation here is more complicated. We expect energy to flow outward, but we also expect energy to flow inward, or equivalently we expect negative energy to flow outward. The reason is that black holes form and we expect the black holes to asymptotically settle to static Shcwarzschild black holes. For this to happen, the ghost matter, which has negative energy, must be expelled from the system. To gain insight, we can look at the energy-momentum tensor.

\begin{comment}

The mass inside the apparent horizons decreases, but there must also be an increase in mass. The reason is that the symmetric massless static wormhole has zero ADM mass. Since figure \ref{fig:EB r symmetric with pulse} indicates that black holes form, and we expect the black holes to asymptotically settle to static Shcwarzschild black holes which have nonzero  and positive ADM mass, there must also be an increase in mass. To gain insight, we can look at the energy-momentum tensor.
\end{comment}

We show the contour diagram for $T_{vv} - T_{uu}$, which gives the net flow of energy-momentum, in figure \ref{fig:EB sigma T m with pulse}(b). Solid gray lines have negative contour values and dashed gray lines have positive contour values. The thick black lines mark where $T_{vv} - T_{uu} = 0$. Focusing on the region $v > u$, we see large portions of spacetime with solid gray lines and hence negative $T_{vv} - T_{uu}$. This indicates outgoing energy-momentum, exactly what we would expect to find for decreasing mass inside the apparent horizons. Near the region where the apparent horizon is close to the the singularity, again for $v > u$, we find dashed lines and hence positive $T_{vv} - T_{uu}$. This indicates ingoing energy-momentum, precisely what is needed for the negative energy ghost matter to flow outward. 

\begin{comment}
We expect the flow of energy-momentum to influence the Misner-Sharp mass, $m(u,v)$, whose contour diagram we show in figure \ref{fig:EB sigma T m with pulse}(c). We include in this diagram the black curves from figure \ref{fig:EB sigma T m with pulse}(b) which mark where $T_{vv} - T_{uu}$ switches sign. We can see in figure \ref{fig:EB sigma T m with pulse}(c) that the black curves cross the contours where the contours make the sharpest change in direction.
\end{comment}

\begin{figure} 
\centering
\includegraphics[width=6.5in]{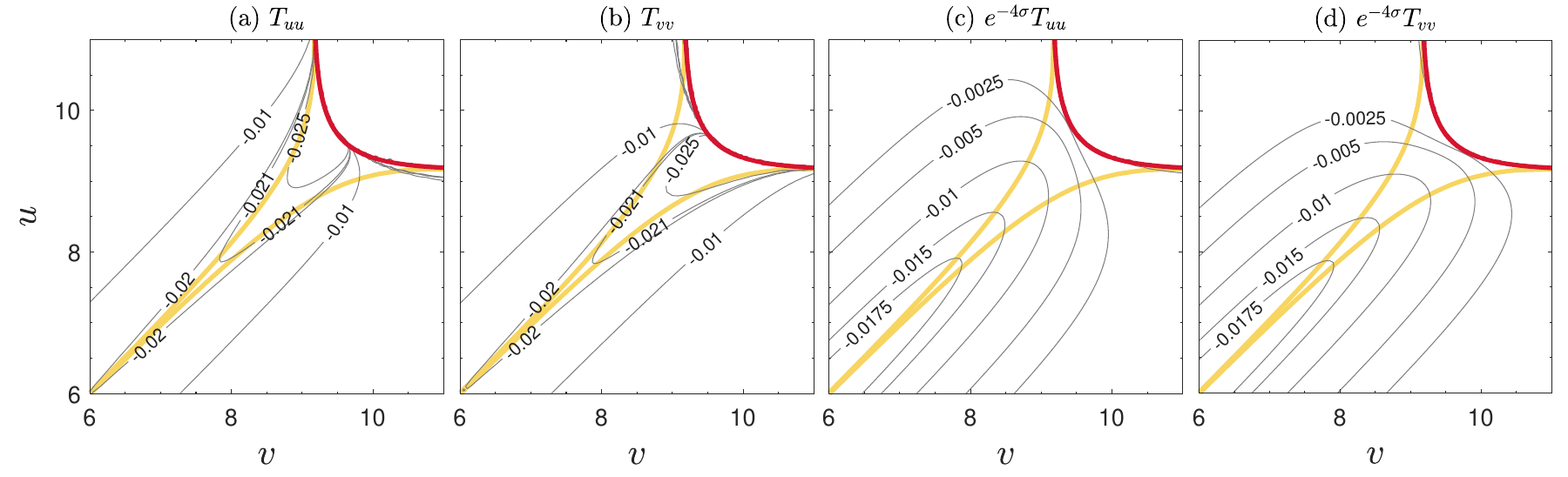} 
\caption{Contours diagrams for (a) outgoing energy-momentum, (b) ingoing energy-momentum, and (c, d) the null energy along null geodesics for the same evolution shown in figure \ref{fig:EB r symmetric with pulse}, with the apparent horizons shown in yellow.}
\label{fig:EB null energy with pulse}
\end{figure}

In figures \ref{fig:EB null energy with pulse}(a) and \ref{fig:EB null energy with pulse}(b), we show $T_{uu}$ and $T_{vv}$ in the region between the apparent horizons. Along a path through the middle of the separated horizons, we find $T_{uu}$ and $T_{vv}$  decreasing, become more negative. Since the wormhole is shrinking, intuitively we would expect the violation of the null energy condition to weaken and for the null energy to head toward zero. Just as with wormhole expansion, we must remember that $T_{uu}$ and $T_{vv}$ do not give the null energy along null geodesics. We plot $e^{-4\sigma}T_{uu}$ and $e^{-4\sigma} T_{vv}$ in figures \ref{fig:EB null energy with pulse}(c) and \ref{fig:EB null energy with pulse}(d), which do increase, heading toward zero, as expected.

\begin{figure} 
\centering
\includegraphics[width=6.5in]{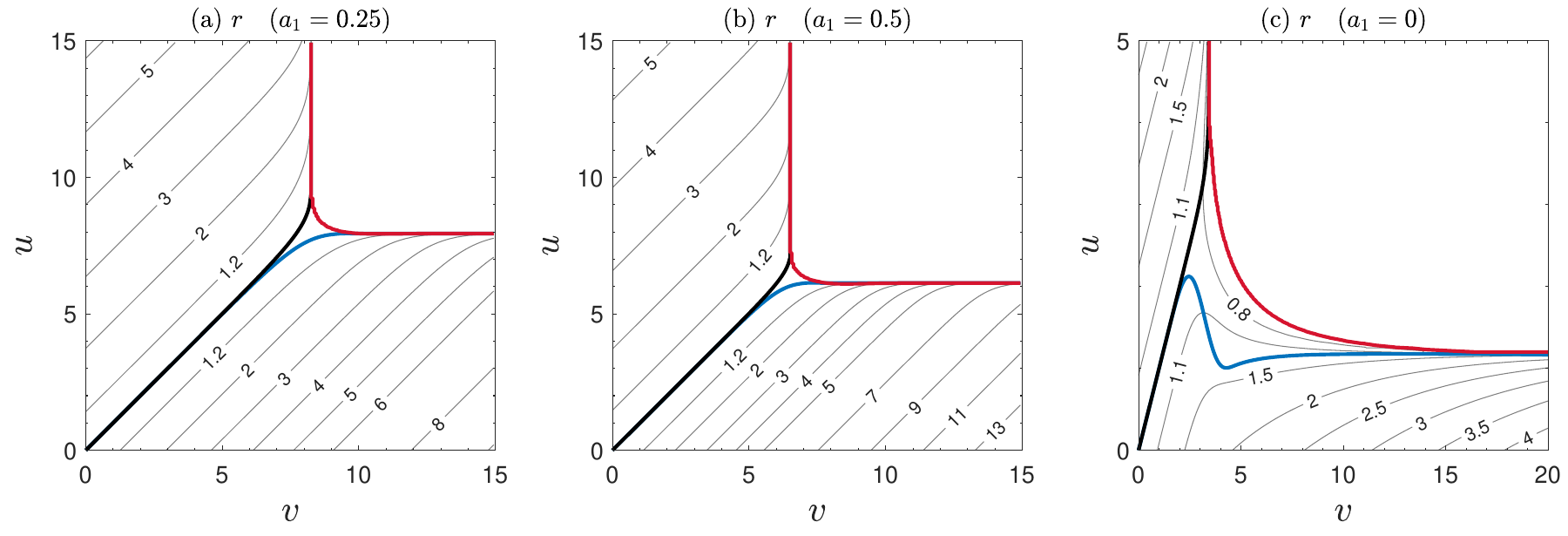} 
\caption{Contour diagrams for the areal radius, $r$. Black and blue curves display apparent horizons $r_{,u} = 0$ and $r_{,v} = 0$ and red curves display central singularities. The initial data for each diagram is an Ellis-Bronnikov wormhole augmented by a scalar field pulse and each diagram shows a collapsing wormhole and the formation of black holes. (a) and (b) use asymmetric wormholes with $a_1 = 0.25$ and 0.5 and the same pulse defined by $A = 0.002$, $v_1 = 1$, and $v_2 = 3$. (c) uses the symmetric wormhole and a much stronger pulse defined by $A = 0.1$, $v_1 = 1$, and $v_2 = 5$. All diagrams are made using grid spacing $\Delta u = \Delta v = 1/400$.}
\label{fig:EB r asymmetric with pulse}
\end{figure}

In figures \ref{fig:EB r asymmetric with pulse}(a) and \ref{fig:EB r asymmetric with pulse}(b), we show contour diagrams for the areal radius for a couple examples of asymmetric massive static solutions perturbed by a pulse of scalar field. The parameters for the pulse are the same as used for the symmetric massless case and gives a relatively weak pulse. In both cases, we can see that the wormhole collapses and black holes form. The asymmetry that can be seen in figures \ref{fig:EB r asymmetric with pulse}(a) and \ref{fig:EB r asymmetric with pulse}(b) comes from the static solution and not from the perturbation. In figure \ref{fig:EB r asymmetric with pulse}(c), we show an example with the symmetric massless static solution and a relatively strong pulse. In this example, the pulse introduces a noticeable asymmetry.

We have investigated the energy-momentum tensors for the examples shown in figure \ref{fig:EB r asymmetric with pulse} and found them to be qualitatively similar to the symmetric massless example detailed above. Specifically, we find large portions of spacetime with negative $T_{vv} - T_{uu}$ which accounts for the decreasing mass between the apparent horizons, while near the region where the apparent horizon is close the singularity, we find positive $T_{vv} - T_{uu}$, which is consistent with the ghost field being expelled and the system asymptotically settling to a Schwarzschild black hole. 

%We also find that the null energy along null geodesics, $e^{-4\sigma}T_{uu}$ and $e^{-4\sigma} T_{vv}$, increases between the apparent horizons, as expected for a collapsing wormhole.

%====================================================================

\section{Quantum corrected Schwarzschild black holes}
\label{sec:QCSBH}

The Schwarzschild black hole is the spherically symmetric vacuum solution in general relativity. In semiclassical gravity, general relativity is extended to include both classical and quantum contributions, where we take the quantum contribution to be a renormalized energy-momentum tensor. The spherically symmetric vacuum solution in semiclassical gravity is the quantum corrected Schwarzschild black hole \cite{Fabbri:2005zn, Ho:2017joh, Ho:2017vgi, Arrechea:2019jgx}. As we'll see, the quantum corrected solution no longer has an event horizon and has a wormhole structure. One side of the wormhole looks very similar to the Schwarzschild black hole and is asymptotically flat. The other side has a null singularity at a finite proper distance from the wormhole throat.

The quantum corrected Schwarzschild black hole, which describes a static wormhole, is sourced by the renormalized energy-momentum tensor. This is quite distinct from the ghost scalar field which sources the Ellis-Bronnikov wormhole. Nonetheless, when dynamically evolved, the quantum corrected Schwarzschild black hole and the Ellis-Bronnikov wormhole are remarkably similar, as we'll see. This leads us to speculate that the results we present in this section, and the results we presented in the previous section for Ellis-Bronnikov wormholes, are typical for wormhole expansion and wormhole collapse.

In section \ref{sec:QCSBH semiclassical gravity}, we discuss semiclassical gravity and the renormalized energy-momentum tensor that we will use. We derive the equations we use to solve for static solutions in section \ref{sec:QCSBH static equations} and the boundary conditions we use to solve the equations in section \ref{sec:QCSBH BC}. In section \ref{sec:QCSBH static solutions}, we review static quantum corrected Schwarzschild black hole solutions. We derive the dynamical equations in section \ref{sec:QCSBH dynamical eqs} and explain how the static solutions are used as initial data in section \ref{sec:QCSBH initial data}. We present results for the expansion and collapse of quantum corrected Schwarzschild black holes in section \ref{sec:QCSBH dynamical evo}.

%---------------------------------------------------------

\subsection{Semiclassical gravity}
\label{sec:QCSBH semiclassical gravity}

Quantum corrected Schwarzschild black holes are spherically symmetric vacuum solutions in semiclassical gravity. In semiclassical gravity, the field equations are still given by (\ref{EFE}), but the energy-momentum tensor is taken to be \cite{Birrell:1982ix}
\begin{equation} \label{semiclassical T}
T_{\mu\nu} = T_{\mu\nu}^{c} + \langle \widehat{T}_{\mu\nu} \rangle,
\end{equation}
where $T_{\mu\nu}^{c}$ is a classical contribution and the semiclassical correction $\langle \widehat{T}_{\mu\nu} \rangle$ is the expectation value of a renormalized energy-momentum tensor operator. When finding static solutions, we set $T_{\mu\nu}^{c} = 0$ since we are interested in vacuum solutions. When dynamically evolving the system, we will at times have a nonzero classical contribution. Just as we did with Ellis-Bronnikov wormholes, we will explicitly perturb the system using a pulse of (regular) scalar field and we will treat the scalar field classically.

The exact form for $\langle \widehat{T}_{\mu\nu} \rangle$ in $3+1$ dimensions is complicated, even for static spherically symmetric spacetimes \cite{Anderson:1994hg}. Since the result contains derivatives higher than second order, it is unclear how to solve the field equations self-consistently and the result tends to be evaluated on fixed spacetime backgrounds. A common alternative is to use the exact renormalized energy-momentum tensor computed in $1+1$ dimensions for the nonangular components in spherically symmetric $3+1$ dimensions \cite{Davies:1976ei, davies, Polyakov:1981rd}. This is known as the Polyakov approximation and leads to a much simpler result, a result that can be solved self-consistently and that is expected to qualitatively agree with the more complicated exact renormalized energy-momentum tensor computed in $3+1$ dimensions.

For simplicity, we focus on computing the renormalized energy-momentum tensor in $1+1$ dimensions for a real massless scalar field. The computation may be performed by using a mode expansion for the scalar field and the point-splitting method as the renormalization procedure \cite{Davies:1976ei, davies, Christensen:1977jc} or by using an effective action and zeta-function regularization as the renormalization procedure \cite{Mottola:2006ew, Mukhanov:2007zz, Hawking:1976ja}. Both methods lead to equivalent results \cite{Barcelo:2011bb}. A useful covariant formula for the renormalized energy-momentum tensor was given in \cite{Barcelo:2011bb}. For the $1+1$-dimensional metric $ds^2 = g_{ab} dx^a dx^b$, where $a,b$ are indices for the nonangular coordinates in $3+1$ dimensions, the renormalized energy-momentum tensor can be computed from
\begin{equation} \label{1+1}
\langle \widehat{T}_{ab}\rangle^{\mathrm{(2D)}}
= \frac{R^{\mathrm{(2D)}}}{48\pi} g_{ab} + \frac{1}{48\pi} \left( \mathcal{A}_{ab} - \frac{1}{2} g_{ab} \mathcal{A}\indices{^c_c} \right),
\end{equation}
where $R^{\mathrm{(2D)}}$ is the $1+1$-dimensional Ricci scalar,
\begin{equation}
\mathcal{A}_{ab} = \frac{4}{|\xi|} \nabla_a \nabla_b |\xi|,
\qquad
|\xi| = \sqrt{-\xi_a \xi^a},
\end{equation}
and $\xi^a$ parameterizes the vacuum state used for the expectation value. Below, we will give explicit forms for $\langle \widehat{T}_{ab}\rangle^{\mathrm{(2D)}}$ for the various coordinate systems we will be using. In the Polyakov approximation, the energy-momentum tensor in $3+1$ dimensions is then given by
\begin{equation} \label{RSET}
\langle \widehat{T}_{ab} \rangle = \frac{1}{4\pi r^2} \langle \widehat{T}_{ab}\rangle^{\mathrm{(2D)}}.
\end{equation}
The multiplicative factor $1/4\pi r^2$ ensures $3+1$-dimensional conservation, $\nabla_\mu \langle \widehat{T}\indices{^\mu_\nu} \rangle = 0$, with $\langle \widehat{T}_{\theta\theta} \rangle = \langle \widehat{T}_{\phi\phi} \rangle = 0$.

A couple of comments are in order. First, instead of a single scalar field used to compute the renormalized energy-momentum tensor, we could assume an arbitrary number. We would then multiply the right hand side of (\ref{RSET}) by the number of scalar fields. For simplicity, we will consider only one. Second, the choice of multiplicative factor in (\ref{RSET}) introduces a divergence as $r\rightarrow 0$. This divergence can be regulated for static solutions by making an alternative choice for the multiplicative factor \cite{Arrechea:2019jgx}. One must then introduce nonzero angular components for $\langle \widehat{T}_{\mu\nu} \rangle$ to maintain conservation. Unfortunately, it does not appear possible to regulate the divergence in time dependent systems and maintain conservation \cite{Parentani:1994ij, Ayal:1997ab}. We will use the multiplicative factor in (\ref{RSET}) and will comment on this divergence when necessary.

In using the Polyakov approximation, we are approximating the $3+1$ renormalized energy-momentum tensor for a massless scalar field using the $1+1$ result. In the context of $3+1$ dimensions, the Polyakov approximation can be understood in terms of two key assumptions for the scalar field. First, only the $s$-wave component of the scalar field's spherical harmonic expansion is retained, with higher order modes neglected. Second, the $s$-wave contribution to the potential is neglected in the equations of motion. A significant advantage of this approach is the resulting simplified expression for the renormalized energy-momentum tensor in (\ref{1+1}), which enables a self-consistent solution of the field equations. Moreover, as long as we avoid the region near $r = 0$, the approximation is expected to qualitatively reproduce the properties of the full, exact renormalized energy-momentum tensor.

%---------------------------------------------------------

\subsection{Static equations}
\label{sec:QCSBH static equations}

Static quantum corrected Schwarzschild black hole solutions are found numerically. Unlike with Ellis-Bronnikov wormholes, we will not be able to find analytical solutions. We parameterize the static metric as
\begin{equation} \label{QCSBH static metric}
ds^2 = e^{2\sigma(x)} (-dt^2 + dx^2) + r^2(x) d\Omega^2.
\end{equation}
Comparing to the general static metric in (\ref{static metric}), we find $\alpha = a = e^\sigma$. When transforming to double null coordinates, we note that $\sigma$ and $r$ in (\ref{QCSBH static metric}) are the same metric functions used in the double null metric in (\ref{metric}).

We can compute the renormalized energy-momentum tensor in the Polykov approximation using (\ref{1+1}). For the metric in (\ref{QCSBH static metric}), the $1+1$-dimensional metric for the nonangular coordinates is $ds^2 = e^{2\sigma}(-dt^2 + dx^2)$. As we'll see, we'll find solutions which asymptotically agree with the classical Schwarzschild solution for $x\rightarrow \infty$. We are therefore looking for static asymptotically flat solutions. The appropriate vacuum for computing the expectation value is then the Boulware vacuum, for which $\xi^a = (1,0)$ for coordinates $(t,x)$. It follows from (\ref{QCSBH static metric}) that the renormalized energy-momentum tensor is
\begin{eqnarray} 
\langle \widehat{T}_{tt} \rangle 
&= - \frac{p}{8 \pi r^2} (\sigma^{\prime \, 2} - 2\sigma'')
\nonumber \\
\langle \widehat{T}_{xx} \rangle 
&= - \frac{p}{8 \pi r^2} \sigma^{\prime \, 2}
\label{static RSET}
\end{eqnarray}
and $\langle \widehat{T}_{\theta\theta} \rangle  = \langle \widehat{T}_{\phi\phi} \rangle = 0 $, where
\begin{equation}
p \equiv \frac{1}{12\pi}.
\end{equation}
We have introduced $p$ in part because it simplifies some equations and in part because it allows us to track which terms originated from the renormalized energy-momentum tensor.

To compute the field equations, we use (\ref{EFE}) and the semiclassical form for the energy-momentum tensor in (\ref{semiclassical T}) with $T_{\mu\nu}^c = 0$. The $tt$, $xx$, and $\theta\theta$ field equations are
\begin{eqnarray} 
\frac{1}{r^2} \left(  2r r'\sigma' -r^{\prime\, 2} -2rr'' + e^{2\sigma}\right)
&= - \frac{p}{r^2} \left(\sigma^{\prime\,2} - 2\sigma'' \right)
\nonumber \\
\frac{1}{r^2} \left( 2rr'\sigma' + r^{\prime\,2} - e^{2\sigma} \right)
&= -\frac{p}{r^2} \sigma^{\prime \, 2} 
\nonumber \\
r e^{-2\sigma} \left( r'' + r \sigma'' \right)
&=  0.
\label{QCSBH EFE}
\end{eqnarray}
Adding the $tt$ and $xx$ field equations together we have an equation with $r''$ and $\sigma''$. Combining this with the $\theta\theta$ field equation, we can eliminate either $\sigma''$ or $r''$. We find
\begin{eqnarray} \label{QCSBC static eqs}
\sigma'' 
&= - \frac{1}{r}\left(2 r' \sigma'
+ \sigma^{\prime\,2}\frac{p}{r} \right)
\left(1 - \frac{p}{r^2} \right)^{-1}
\nonumber \\
r'' 
&= \left(2 r' \sigma'
+ \sigma^{\prime\,2}\frac{p}{r} \right)
\left(1 - \frac{p}{r^2} \right)^{-1}.
\end{eqnarray}
We shall use these equations to numerically solve for static solutions. Doing so requires boundary conditions, which we determine in the next subsection. Note that these equations are divergent when $r\rightarrow \sqrt{p}$. This divergence is a consequence of the multiplicative factor in (\ref{RSET}). For convenience, we will focus on static solutions which do not approach this divergence because the areal radius will always be greater than $\sqrt{p}$. As such, at least for static solutions, this divergence will not be a concern.

There remains the $xx$ field equation, which is the middle equation in (\ref{QCSBH EFE}). Since the equation is quadratic in $\sigma'$, we can straightforwardly solve for it,
\begin{equation} \label{sigma prime eq}
\sigma' =  - \frac{r r'}{p} \left[
1 \pm \sqrt{
1 + \frac{ p }{(r r')^2} \left( e^{2\sigma} - r^{\prime \, 2} \right)
}\right].
\end{equation}
This equation offers some interesting insight into how the renormalized energy-momentum tensor affects solutions. The renormalized energy-momentum tensor vanishes in the classical limit, which can be parametrized by $p\rightarrow 0$. Expanding (\ref{sigma prime eq}) around small $p$ gives
\begin{equation}
\sigma' = - (1 \pm 1) \frac{r r'}{p}
\mp \frac{e^{2\sigma} - r^{\prime \, 2}}{2 r r'} 
+ O(p).
\end{equation}
We find that the classical limit $p\rightarrow 0$ exists for the lower sign in (\ref{sigma prime eq}). When using the lower sign, we can interpret the renormalized energy-momentum tensor as a quantum perturbation to the classical equations. The classical limit does not exist for the upper sign in (\ref{sigma prime eq}). In this case, we interpret the renormalized energy-momentum tensor as a nonperturbative correction. After we solve (\ref{QCSBC static eqs}), we can compare the solution against (\ref{sigma prime eq}) and determine whether the solution is perturbative or nonperturbative.

%---------------------------------------------------------

\subsection{Boundary conditions}
\label{sec:QCSBH BC}

In the previous subsection, we derived the field equations in (\ref{QCSBC static eqs}). These field equations parameterize the metric functions $\sigma$ and $r$ in terms of the radial coordinate $x$. This parameterization is helpful for transforming solutions to double null coordinates, as we saw in section \ref{sec:initial data}. However, the asymptotic behavior of the solutions in the large $x$ limit is somewhat complicated, making the study of the outer boundary conditions using these variables less transparent.

It is possible to decouple the field equations in (\ref{QCSBC static eqs}) by treating $r$ as the independent variable and to find a single field equation for $\sigma(r)$. This equation is easier to analyze in the large $r$ limit and will help us find outer boundary conditions. In decoupling the field equations, we indicate an $r$ derivative with an overdot, $\dot{\sigma} = \partial_r \sigma$, and the relationship between $x$ and $r$ derivatives is
\begin{equation} \label{deriv convert}
\sigma' = r' \dot{\sigma},
\qquad
\sigma''= r'' \dot{\sigma} + r^{\prime\,2} \ddot{\sigma}.
\end{equation}
Converting the derivatives in (\ref{QCSBC static eqs}), the resulting equations can be combined to eliminate $r''$. We find
\begin{equation} \label{sigma(r) ode}
0 = \ddot{\sigma}  \left(1  - \frac{ p }{r^2} \right)
+ \dot{\sigma}^3 \frac{ p}{r} 
+ \dot{\sigma}^2 \left( 2 + \frac{ p }{r^2} \right)
+ \dot{\sigma}\frac{2}{r},
\end{equation}
which is the desired formula.

We cannot solve for static wormhole solutions using (\ref{sigma(r) ode}) in its present form. The reason is that the areal radius $r$ has a minimum. Outside this minimum, $\sigma$ is double valued for a single value of $r$. It is possible to invert (\ref{sigma(r) ode}), deriving an equation for $r(\sigma)$, with $\sigma$ the independent variable. One can then solve for static wormhole solutions, as was done in \cite{Fabbri:2005zn}. Having said this, we reiterate that we prefer to solve for static solutions using (\ref{QCSBC static eqs}) since the solutions can more easily be transformed to double null coordinates.

Although we cannot solve for the full wormhole solution using (\ref{sigma(r) ode}), we can solve for the solution on one side of the wormhole throat. In particular, we can analytically solve for the asymptotic solution at large $r$. The value in solving for the asymptotic solution using (\ref{sigma(r) ode}) instead of (\ref{QCSBC static eqs}) is that the solution has the simple form
\begin{equation} \label{sigma expansion}
\sigma = \sigma_0 + \frac{\sigma_1}{r} + \frac{\sigma_2}{r^2} + \frac{\sigma_3}{r^3} + \cdots.
\end{equation}
We can set $\sigma_0 = 0$ since we will look for asymptotically flat solutions. Plugging (\ref{sigma expansion}) into (\ref{sigma(r) ode}), we can solve for the expansion coefficients, finding
\begin{equation} \label{sigma expansion 2}
\sigma = - \frac{M}{r} - \frac{M^2}{r^2} - \frac{4M^3 + pM}{3r^3} + \cdots,
\end{equation}
where $M$ is an undetermined constant. Importantly, we find that the renormalized energy-momentum tensor enters the large $r$ solution at order $r^{-3}$, as indicated by the presence of $p$. In other words, at sufficiently large $r$ the static solution agrees with the classical solution, i.e.~the Schwarzschild solution. Having established this, we can match the Schwarzschild solution to the expansion in (\ref{sigma expansion 2}) and show that $M$ is the Schwarzschild mass parameter. We have shown that the classical Schwarzschild solution can be used as our outer boundary condition.

To find the classical Schwarzschild solution, we can analytically solve (\ref{sigma(r) ode}) for $\sigma(r)$ after setting $p = 0$. We can use the result in the middle equation of (\ref{QCSBH EFE}), after setting $p =0$ and converting derivatives using (\ref{deriv convert}), and analytically solve for $x(r)$. The classical Schwarzschild solution is
\begin{equation} \label{classical Schwarzschild}
\sigma = \frac{1}{2} \ln \left(1 - \frac{2M}{r} \right),
\qquad
x = r + 2M \ln \left( \frac{r}{2M} - 1 \right),
\end{equation}
where we have set integration constants to their conventional values. In the context of the Schwarzschild solution, $x$ is known as the tortoise coordinate. Since (\ref{QCSBC static eqs}) is a system of second order ODEs, we will also need first derivatives. Taking $x$ derivatives,
\begin{equation} \label{classical Schwarzschild 2}
\sigma' = \frac{M}{r^2},
\qquad
r' = 1- \frac{2M}{r}.
\end{equation}

%---------------------------------------------------------

\subsection{Static solutions}
\label{sec:QCSBH static solutions}

The equations for finding static solutions and the boundary conditions for solving the equations were derived in the previous two subsections. Our method for finding static quantum corrected Schwarzschild black hole solutions is now clear. We choose a value for $M$, which is the classical Schwarzschild mass and the only free parameter. We then choose a large value for $r$ and compute the corresponding values of $\sigma, x, \sigma', r'$ using (\ref{classical Schwarzschild}) and (\ref{classical Schwarzschild 2}), which are the values at the outer boundary of the numerical solution. We then solve (\ref{QCSBC static eqs}) numerically by integrating inward. Solutions will be independent of the chosen value for $r$ as long as it is sufficiently large.

\begin{figure} 
\centering
\includegraphics[width=6.5in]{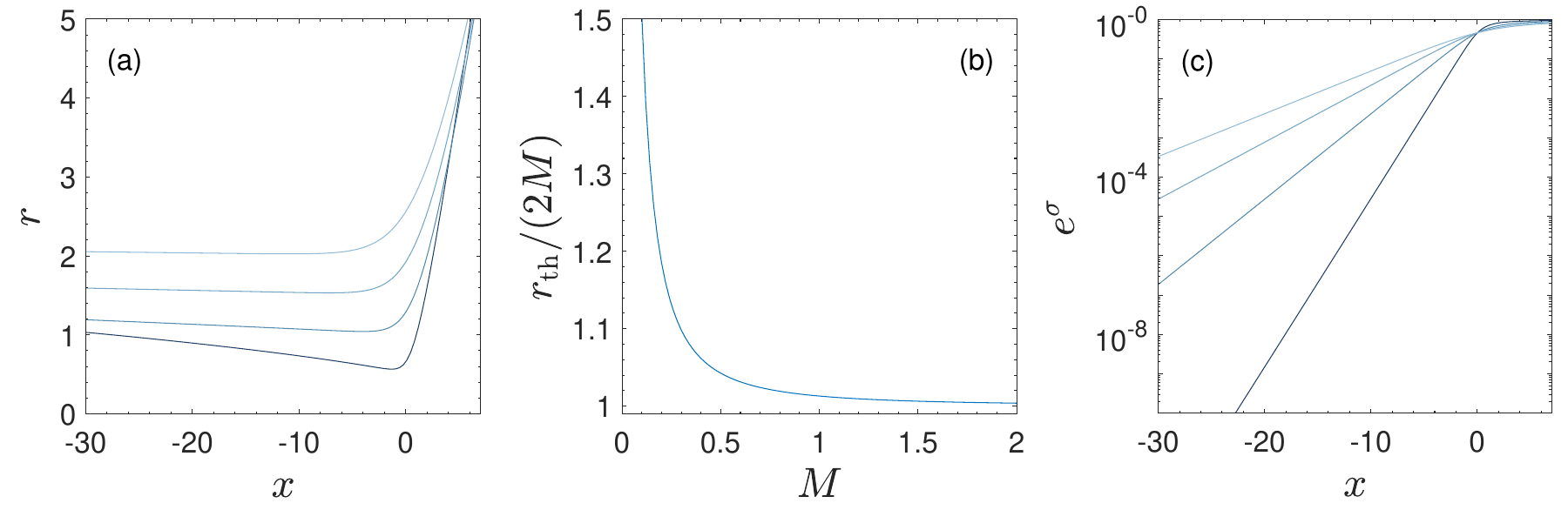} 
\caption{These diagrams display static quantum corrected Schwarzschild black hole solutions, which are parameterized in terms of the classical Schwarzschild mass, $M$. (a) The areal radius, $r$, as a function of the radial coordinate, $x$, for (from darkest curve to lightest) $M = 0.25$, 0.5, 0.75, and 1. (b) The ratio of the wormhole throat radius, $r_\mathrm{th}$, to the Schwarzschild radius as a function of $M$. (c) The metric component $e^\sigma$ as a function of $x$ for the same solutions shown in (a).}
\label{fig:QCSBC static solutions}
\end{figure}

In figure \ref{fig:QCSBC static solutions}(a), we show the areal radius, $r(x)$, for $M = 0.25$, 0.5, 0.75, and 1. A minimum is easily seen for smaller values of $M$. Though less obvious, a minimum persists as $M$ increases. Each solution therefore has a wormhole structure. As $M$ increases, the wormhole throat radius $r_\mathrm{th}$, which is the minimum value of $r$, also increases and the solution becomes more asymmetric.

The classical Schwarzschild radius is equal to $2M$. We show the ratio of the wormhole throat radius to the Schwarzschild radius as function of $M$ in figure \ref{fig:QCSBC static solutions}(b). We find that the wormhole throat radius is always larger, but asymptotically approaches the Schwarzschild radius as $M$ increases.

In figure \ref{fig:QCSBC static solutions}(c), we show the metric component $e^\sigma$ for each of the solutions shown in figure \ref{fig:QCSBC static solutions}(a). Since $e^\sigma$ is nonzero, there is no event horizon in the plotted region. One can show that the classical event horizon has disappeared completely in the quantum corrected solution \cite{Fabbri:2005zn, Ho:2017joh, Ho:2017vgi, Arrechea:2019jgx}. The renormalized-energy momentum tensor has removed the event horizon and replaced it with a wormhole structure. We do find that $e^\sigma$ is small. As a consequence, observations of the static system will closely resemble observations of a system with a horizon. By construction, the static solution asymptotically agrees with the Schwarzschild solution at large positive $x$.  Putting these two facts together, when observed from large $x$, it would be difficult to distinguish this system from a Schwarzschild black hole, even though it is a horizonless wormhole.

It was shown in \cite{Fabbri:2005zn} that a null curvature singularity exists in the limit $x\rightarrow -\infty$ and that this singularity is located a finite proper distance from the wormhole throat. As such, only one side of the wormhole is asymptotically flat. It may therefore be more appropriate to refer to this system as having a wormhole structure instead of referring to this system as a wormhole. We shall continue to refer to this system as a wormhole out of convenience.

\begin{figure} 
\centering
\includegraphics[width=4.25in]{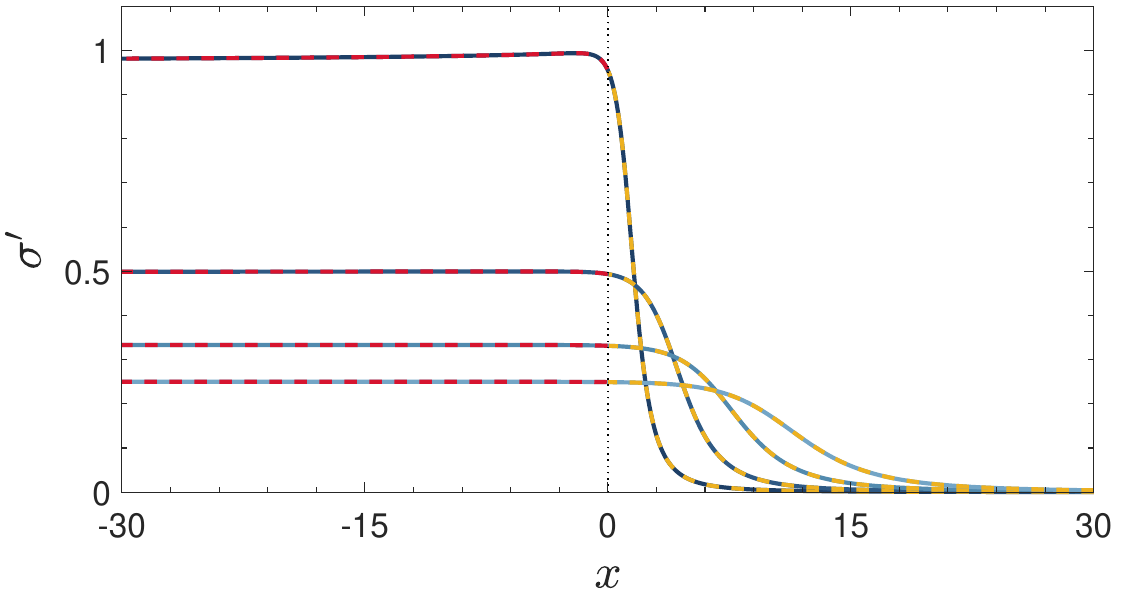} 
\caption{The blue curves plot $\sigma'$ for the same solutions shown in figure \ref{fig:QCSBC static solutions}(a). The dashed yellow curves are from (\ref{sigma prime eq}) with the lower sign and the dashed red curves are from (\ref{sigma prime eq}) with the upper sign. All curves have been shifted so that the wormhole throat is located at $x = 0$, which is marked by the vertical dotted line.}
\label{fig:QCSBC (non)perturbative}
\end{figure}

We show $\sigma'$ in figure \ref{fig:QCSBC (non)perturbative} for each of the solutions shown in figure \ref{fig:QCSBC static solutions}(a) as the solid blue curves. We have shifted each curve so that the wormhole throat is at $x = 0$, which is indicated by the vertical dotted line. On top of each curve, we have plotted (\ref{sigma prime eq}) with the lower sign as dashed yellow and (\ref{sigma prime eq}) with the upper sign as dashed red. We have only plotted (\ref{sigma prime eq}) with either sign over the region where it agrees with the numerical solution (the blue curves). We recall that when using the lower sign in (\ref{sigma prime eq}), the renormalized energy-momentum tensor can be interpreted as a perturbation to the classical result and when using the upper sign, it can be interpreted as a nonperturbative correction. We find from figure \ref{fig:QCSBC (non)perturbative} that the renormalized energy-momentum tensor can be interpreted as a perturbation at large positive $x$. This is expected, since we compute the numerical solutions by using the classical result as the outer boundary condition. That it can be interpreted as a perturbation persists up until the wormhole throat. Interestingly, precisely at the wormhole throat the solution switches continuously from perturbative to nonperturbative. The solution then remains nonperturbative as $x \rightarrow -\infty$. Returning to figure \ref{fig:QCSBC static solutions}(a), coming in from large positive $x$, the areal radius decreases. The use of (\ref{sigma prime eq}) has shown us that the areal radius reaches a minimum and then turns up and starts increasing because of a nonperturbative effect introduced by the energy-momentum tensor. As such, the wormhole throat cannot be approximated with a perturbative expansion \cite{Fabbri:2005zn}.

\begin{comment}
Regions to the right and to the left of the wormhole throat make up the two sides of the wormhole. The region to the right asymptotically agrees with the Schwarzschild solution and is asymptotically flat. If this wormhole were to exist, this is the region where we would reside. We will refer to this region as ``outside" the wormhole. We will refer to the opposite side of the wormhole, where the null singularity resides, as ``inside" the wormhole. We find these to be convenient labels. 
\end{comment}

%---------------------------------------------------------

\subsection{Dynamical equations}
\label{sec:QCSBH dynamical eqs}

To dynamically evolve the quantum corrected Schwarzschild black hole, we use double null coordinates, as we discussed in section \ref{sec:double null}. We can compute the renormalized energy-momentum tensor in double null coordinates using (\ref{1+1}). For the double null metric in (\ref{metric}), the $1+1$-dimensional metric for the nonangular coordinates is $ds^2 = -e^{2\sigma}dudv$. We must also supply the vector $\xi^a$. In the static case, using coordinates $(t,x)$, we used $\xi^a = (1,0)$. Transforming to double null coordinates $(u,v)$, this becomes $\xi^a = (1,1)$. It follows from (\ref{1+1}) that the renormalized energy-momentum tensor is
\begin{eqnarray} \label{RSET uv}
\langle\widehat{T}_{uu}\rangle 
&= 
\frac{p}{4\pi r^2}
(\sigma_{,uu} - \sigma_{,u}^2)
\nonumber \\
\langle\widehat{T}_{vv}\rangle 
&= 
\frac{p}{4\pi r^2}
(\sigma_{,vv} - \sigma_{,v}^2)
\nonumber \\
\langle  \widehat{T}_{uv}\rangle
&= 
- \frac{p}{4\pi r^2}
\sigma_{,uv}
\end{eqnarray}
and $\langle \widehat{T}_{\theta\theta} \rangle  = \langle \widehat{T}_{\phi\phi} \rangle = 0$ \cite{Parentani:1994ij, Ayal:1997ab, Sorkin:2001hf, Hong:2008mw, Hwang:2010im}.

We may use the field equations in (\ref{constraint eqs}) and (\ref{sigma r evo}), but with the energy-momentum tensor given by (\ref{semiclassical T}) and (\ref{RSET uv}). Plugging in for the energy-momentum tensor, the constraint equations become
\begin{eqnarray} \label{constraint eqs with RSET}
r_{,uu} 
&= 2 \sigma_{,u} r_{,u} - 4\pi r \left[ T_{uu}^c 
+ \frac{p}{4\pi r^2}
(\sigma_{,uu} - \sigma_{,u}^2) \right]
\nonumber \\
r_{,vv} 
&= 2 \sigma_{,v} r_{,v} - 4\pi r \left[ T_{vv}^c
+ \frac{p}{4\pi r^2}
(\sigma_{,vv} - \sigma_{,v}^2) \right],
\end{eqnarray}
and the evolution equations become
\begin{eqnarray} \label{sigma r evo with RSET}
\sigma_{,uv} 
&= \frac{1}{4r^2} \left[4 r_{,u} r_{,v} + e^{2\sigma} 
- 8\pi \left( 2 r^2 T_{uv}^c +  e^{2\sigma}   T_{\theta\theta}^c \right)
\right]
\left( 1 - \frac{p}{r^2} \right)^{-1}
\\\nonumber 
r_{,uv} 
&= - \frac{1}{4r} \left[4 r_{,u} r_{,v} + e^{2\sigma} 
- 16\pi r^2 \left( T_{uv}^c
- \frac{p}{4\pi r^2}
\sigma_{,uv} \right) \right].
\end{eqnarray}
Note that the $\sigma_{,uv}$ equation is divergent when $r\rightarrow \sqrt{p}$, as is the $r_{,uv}$ equation since it depends on $\sigma_{,uv}$. This divergence is a consequence of the multiplicative factor in (\ref{RSET}). It is also present in the static equations in (\ref{QCSBC static eqs}), but did not play a role in our static solutions because we focused on solutions for which the areal radius was always greater than $\sqrt{p}$. In dynamically evolving the system, if the wormhole collapses, then the areal radius will decrease and reach $\sqrt{p}$, as we'll see. A common interpretation for this divergence is to regard it as corresponding to the central singularity, which semiclassical effects have shifted from $r = 0$ to $r = \sqrt{p}$ \cite{Sorkin:2001hf, Fabbri:2005mw, Hong:2008mw, Hwang:2010im}. With this interpretation, the divergence will not be a cause for concern as long as we make sure that any length scale in the system, such as horizon radii, are large compared to $\sqrt{p}$.

We will use a real massless scalar field, $\chi$, to explicitly perturb the system, just as we did with Ellis-Bronnikov wormholes. $\chi$ will be the only matter in the system and the Lagrangian for the matter sector is
\begin{equation} 
\mathcal{L} = 
- \frac{1}{2} (\nabla_\mu \chi)(\nabla^\mu \chi),
\end{equation}
which we minimally couple to gravity,  $\mathcal{L} \rightarrow \sqrt{-\det(g_{\mu\nu})} \, \mathcal{L}$. We previously worked out the necessary equations in double null coordinates during our study of Ellis-Bronnikov wormholes.  The equation of motion is the massless Klein-Gordon equation, $\nabla_\mu \nabla^\mu \chi = 0$. For the double null metric in (\ref{metric}), the equation of motion is given in the second equation in (\ref{phi chi eom}). The only contribution to the classical energy-momentum tensor is the energy-momentum tensor for $\chi$, $T_{\mu\nu}^c = T_{\mu\nu}^\chi$, where $T_{\mu\nu}^\chi$ is given in (\ref{EB T chi}).

%---------------------------------------------------------

\subsection{Initial data}
\label{sec:QCSBH initial data}

To construct initial data from a static wormhole solution, we follow the methods described in section \ref{sec:initial data}. These methods make use of the new radial coordinate, $y$, defined by the integral in (\ref{y from x}). For the static metric we are using in (\ref{QCSBH static metric}), $y$ is just a linear shift of $x$. We choose the integration constants such that the wormhole throat is at the origin of the computational domain,
\begin{equation} \label{QCSBH x to y}
y = y_0 + x - x_\mathrm{th},
\end{equation}
where $y_0 = (v_i - u_i)/2$ and $x_\mathrm{th}$ is the location of the wormhole throat. $y$ can be transformed to double null coordinates using (\ref{t y from u v}). Along the initial hypersurfaces, we have
\begin{equation} \label{u v to x}
x(u,v_i) = x_\mathrm{th} - \frac{1}{2}(u - u_i),
\qquad
x(u_i,v) = x_\mathrm{th} + \frac{1}{2}(v - v_i).
\end{equation}
For any point on an initial hypersurface, we have the corresponding value of $x$ from (\ref{u v to x}), from which we can compute the value of any field from the static solution.

\begin{figure} 
\centering
\includegraphics[width=6.5in]{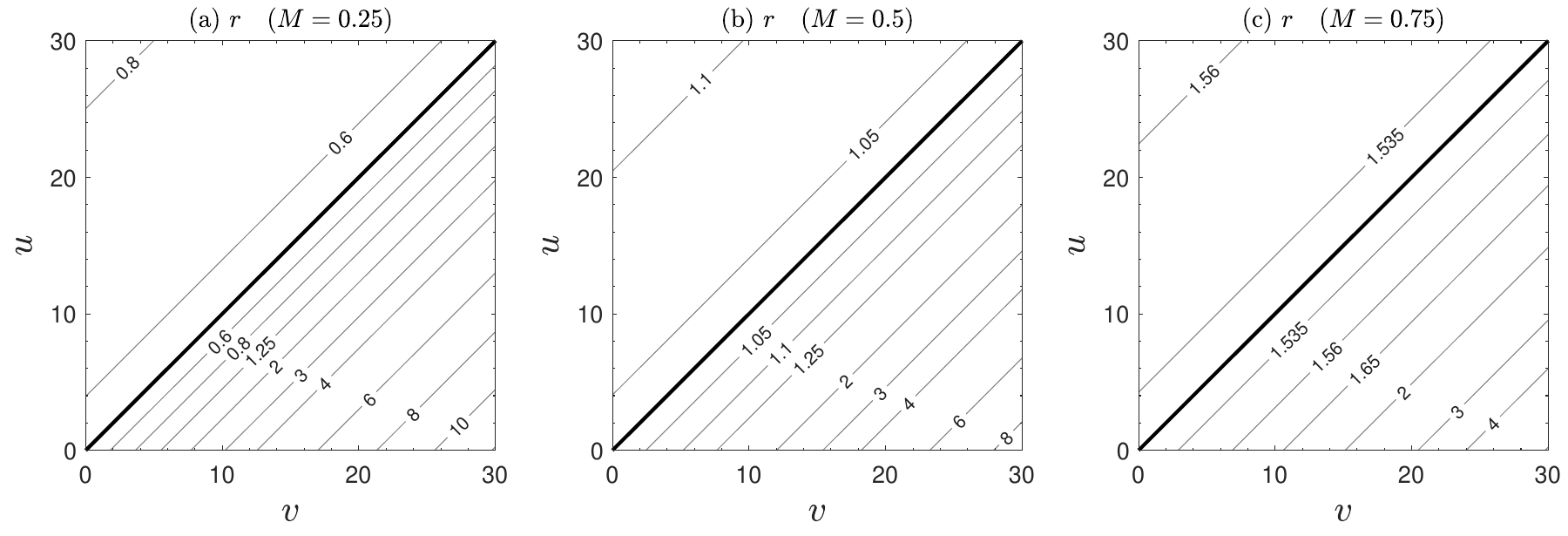} 
\caption{Contour diagrams for the areal radius, $r$, in double null coordinates for static quantum corrected Schwarzschild black holes defined by (a) $M = 0.25$, (b) 0.5, and (c) 0.75. In each diagram, the gray lines are contours with values of $r$ as labeled and the thick black lines mark the wormhole throat with radii (a) $r_\mathrm{th} = 0.566$, (b) 1.043, and (c) 1.532.}
\label{fig:static QCSBH uv}
\end{figure}

When discussing the initial data for the dynamical evolution of the Ellis-Bronnikov wormhole in section \ref{sec:EB initial data}, we presented contour diagrams for the areal radius for static solutions. We did this to give a sense for the diagrams in anticipation of their usefulness for understanding the dynamical evolution of the system. We can do the same here. Figure \ref{fig:static QCSBH uv} displays the contour diagram for $r$ for various static solutions and is analogous to figure \ref{fig:QCSBC static solutions}(a), but in double null coordinates. The thick black line in each plot is the contour for $r = r_\mathrm{th}$, where $r_\mathrm{th} = 0.566$, $1.043$, and $1.532$ in figures  \ref{fig:static QCSBH uv}(a),  \ref{fig:static QCSBH uv}(b), and  \ref{fig:static QCSBH uv}(c). The two sides of the thick black line plot the two sides of the wormhole and the asymmetry of each wormhole is apparent. Above the thick black line, far fewer contour lines are plotted than below. We can see why this is the case by looking at figure \ref{fig:QCSBC static solutions}(a):~To the left of the wormhole throat, the curves have a small slope and so the curves pass through a small range of $r$ values for a comparatively large range of $x$ values. It is this region that is plotted above the thick black lines in figure \ref{fig:static QCSBH uv}. We remind that lines of constant $v-u$ are lines of constant $y$ and hence lines of constant $x$, as follows from (\ref{t y from u v}). For this reason, static systems have straight contour lines with unit slope.

For the dynamical evolution, the initial data is placed on the $u = u_i$ and $v = v_i$ hypersurfaces, which are the axes in figure \ref{fig:static QCSBH uv}. For initial data, we can use a static solution or we can augment a static solution with a pulse of scalar field, which acts as an explicit perturbation. Similarly to what we did with Ellis-Bronnikov wormholes, when including an explicit perturbation we will use a single pulse on the $u = u_i$ hypersurface given by (\ref{chi pulse 1}) and (\ref{chi pulse 2}) (but with $x_0$ set to $1$).

When including a pulse, the energy-momentum tensor has a nonzero classical contribution from $T_{\mu\nu}^\chi$ along the $u = u_i$ hypersurface. The initial data on the $v = v_i$ hypersurface is therefore unchanged from the static solution, but $\sigma$ and $r$ on the $u = u_i$ hypersurface may change and must be computed. We listed two standard options for computing $\sigma$ or $r$ in our discussion of initial data for Ellis-Bronnikov wormholes in section \ref{sec:EB initial data}. One of these options was to keep $r(u_i, v)$ unchanged and to compute $\sigma(u_i,v)$ using the relevant constraint equation. This was not the option we made use of, but we confirmed that it worked perfectly well for Ellis-Bronnikov wormholes. However, we do not find it works well here, in that we had trouble numerically solving the constraint equation. The relevant constraint equation is the second equation in (\ref{constraint eqs with RSET}). We suspect the issue is that the constraint equation must be solved for $\sigma_{,vv}$, but $\sigma_{,vv}$ comes from the renormalized energy-momentum tensor. We therefore use the same option that we used for Ellis-Bronnikov wormholes:~We keep $\sigma(u_i,v)$ unchanged from the static solution and compute $r(u_i,v)$ using the second constraint equation in (\ref{constraint eqs with RSET}). 

In solving the constraint equation on the $u = u_i$ initial hypersurface, we need $\sigma_{,v}(u_i,v)$ and $\sigma_{,vv}(u_i,v)$, which are unchanged from the static solution, and $r(u_i, v_i)$ and $r_{,v}(u_i, v_i)$ at the origin, which are also unchanged from the static solution since we will only consider pulses outside the origin. The static solution is parameterized in terms of $x$, which is related to double null coordinates through (\ref{QCSBH x to y}) and (\ref{t y from u v}). We have,
\begin{equation}
\sigma_{,v} = \frac{1}{2} \sigma',
\qquad
\sigma_{,vv} = \frac{1}{4} \sigma'',
\qquad
r_{,v} = \frac{1}{2} r',
\end{equation}
where $\sigma'$ and $r'$ are obtained from the numerical solution and $\sigma''$ is given in (\ref{QCSBC static eqs}). We can now integrate the constraint equation outward from $v = v_i$ to obtain the updated $r(u_i,v)$.

%---------------------------------------------------------

\subsection{Dynamical evolution}
\label{sec:QCSBH dynamical evo}

We now present results for the dynamical evolution of quantum corrected Schwarzschild black holes \cite{Kain:2025lxd}. We analyze these results similarly to how we analyzed Ellis-Bronnikov wormholes in section \ref{sec:EB dynamical evo}. Some explanations in this section will therefore be briefer.  We will find that the dynamical evolution of the two types of wormholes share many similarities, which we will discuss in the next section. We begin with examples of wormhole expansion before presenting examples of wormhole collapse.

%+++++++++++++++++++++++++++++++++++++++++++++++++++++++++

\subsubsection{Wormhole expansion}
\label{sec:QCSBH expansion}

\begin{figure} 
\centering
\includegraphics[width=6.5in]{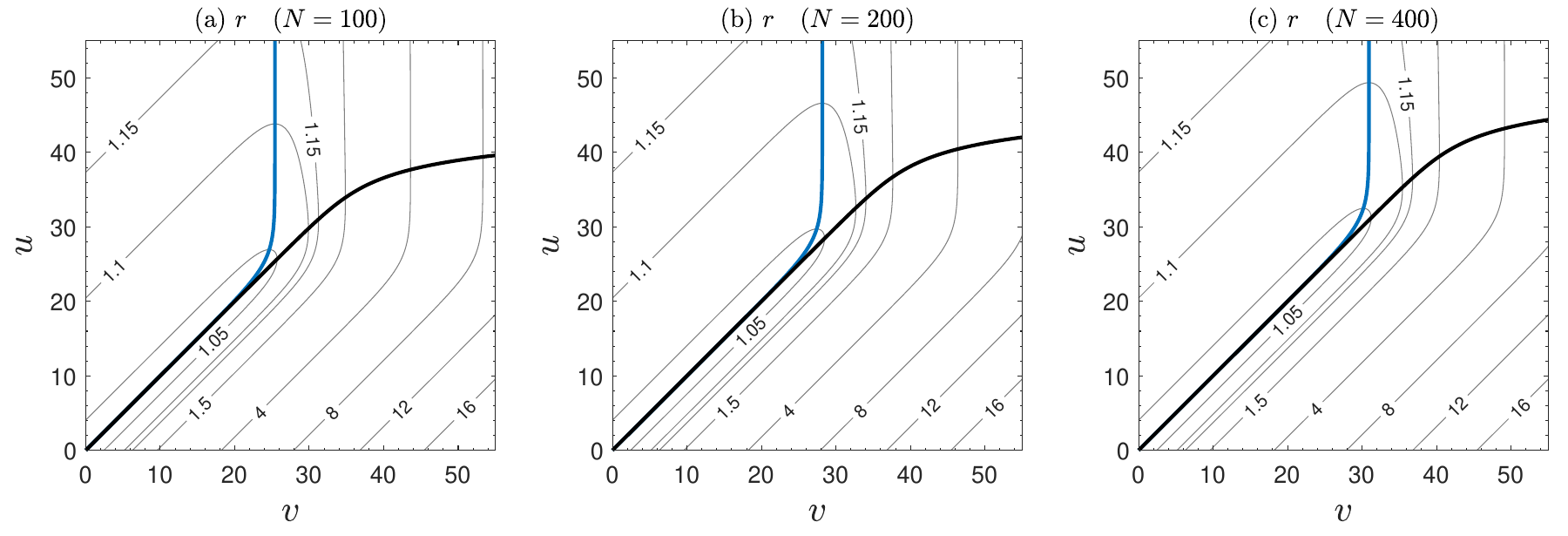} 
\caption{Contour diagrams for the areal radius, $r$, for the dynamical evolution of a quantum corrected Schwarzschild black hole with $M = 0.5$. Each diagram displays an expanding wormhole. The black curves display the apparent horizon $r_{,u} = 0$ and the blue curves display the apparent horizon $r_{,v} = 0$. Each diagram uses the same initial data, but evolves the system with grid spacing $\Delta u = \Delta v = 1/N$ with (a) $N = 100$, (b) 200, and (c) 400. As the grid spacing decreases, the evolution maintains the static configuration longer. This is the expected behavior for a unstable static solution.}
\label{fig:QCSBH r symmetric no pulse}
\end{figure}

In figure \ref{fig:QCSBH r symmetric no pulse}(a), we show a contour diagram for the areal radius for the dynamical evolution of a static solution with $M = 0.5$. As a reminder, the static solution is placed on the initial hypersurfaces, which are the axes in the diagram, the gray lines are the contour lines for $r$, with the values for $r$ as labeled, and the thick black and blue lines are apparent horizons defined by $r_{,u} = 0$ and $r_{,v} = 0$, respectively. Comparing figure \ref{fig:QCSBH r symmetric no pulse}(a) to the static solution in figure \ref{fig:static QCSBH uv}(b), it is clear that the diagrams are different, indicating that the system in figure \ref{fig:QCSBH r symmetric no pulse}(a) evolves away from the static configuration.

To understand the evolution, we start in the bottom left corner. Moving away from this corner, the diagram is initially identical to figure \ref{fig:static QCSBH uv}(b), indicating that the system is maintaining the static configuration. The apparent horizons separate at around $u \approx 25$, at which point the system evolves away from the static configuration. In between the separated apparent horizons, moving in a timelike direction moves across increasing radii and we find that the wormhole is expanding, which appears to continue indefinitely. The blue curve evolves to being vertical, signifying a cosmological horizon at $v \approx 25$. If we were to extend the diagram to larger values of $v$, we would find that the black curve asymptotically approaches horizontal, signifying a cosmological horizon at $u \approx 45$.

The only perturbation in the dynamical evolution shown in figure \ref{fig:QCSBH r symmetric no pulse}(a) is the discretization error inherent to the code, since we did not include a scalar field pulse. We can decrease the strength of this perturbation by increasing the number of grid points used in the numerical computation. Figure \ref{fig:QCSBH r symmetric no pulse}(a) is made with $N = 100$ grid points per unit interval, with the spacing between grid points equal to $\Delta u = \Delta v = 1/N$. Figures \ref{fig:QCSBH r symmetric no pulse}(b) and \ref{fig:QCSBH r symmetric no pulse}(c) use the same initial data, but evolve the system using $N = 200$ and $400$ grid points per unit interval. As the strength of perturbation is decreased, we find that the system is able to maintain the static configuration longer, before eventually evolving away from it. We have therefore shown that the static solution is nonlinearly unstable with respect to time dependent perturbations. Unlike with the Ellis-Bronnikov wormhole, a linear stability analysis has not yet been performed for the static quantum corrected Schwarzschild black hole, although these results hint at it also being linearly unstable.

\begin{figure} 
\centering
\includegraphics[width=6.5in]{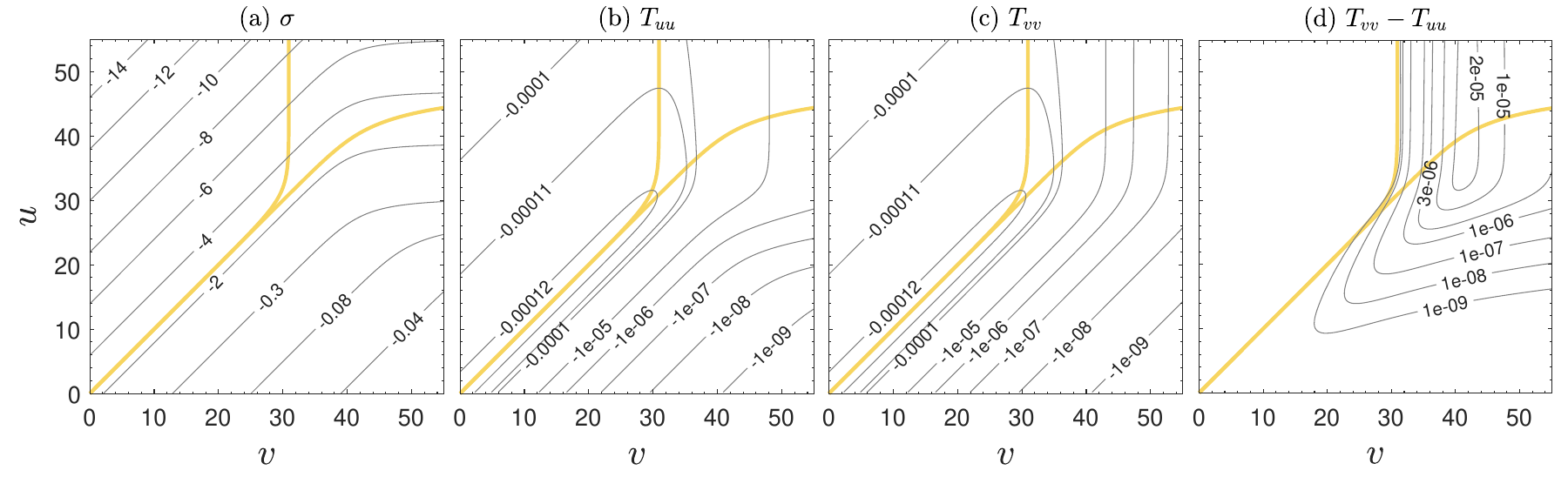} 
\caption{Each diagram is for the same dynamical evolution shown in figure \ref{fig:QCSBH r symmetric no pulse}(c). The yellow curves are the apparent horizons from figure \ref{fig:QCSBH r symmetric no pulse}(c). The diagrams display contours for (a) the metric field $\sigma$, (b) the outgoing energy-momentum $T_{uu}$, (c) the ingoing energy-momentum $T_{vv}$, and (d) the net flow of energy momentum, $T_{vv} - T_{uu}$. In (d), the regions left blank are difficult to compute numerically due to cancellations which are close to zero.}
\label{fig:QCSBH sigma T no pulse}
\end{figure}

The diagrams in figure \ref{fig:QCSBH sigma T no pulse} are for the same evolution shown in figure \ref{fig:QCSBH r symmetric no pulse}(c). The apparent horizons are included in yellow for convenience. We show the contour diagram for the metric field $\sigma$ in figure \ref{fig:QCSBH sigma T no pulse}(a). 
%Following the evolution from the bottom left corner to the top right corner, after the apparent horizons separate, $\sigma$ mostly decreases. 

Once the system evolves away from the static configuration and the wormhole is expanding, the apparent horizons evolve along increasing radii, as can be seen in figure \ref{fig:QCSBH r symmetric no pulse}. We therefore find that the mass inside the apparent horizons increases during expansion. For the mass to increase, we expect energy to flow inward. We show contour diagrams for outgoing energy-momentum $T_{uu}$ in figure \ref{fig:QCSBH sigma T no pulse}(b), ingoing energy-momentum $T_{vv}$ in figure \ref{fig:QCSBH sigma T no pulse}(c), and the net flow of energy-momentum, $T_{vv} - T_{uu}$, in figure \ref{fig:QCSBH sigma T no pulse}(d). In the region near future null infinity, the net flow is positive, indicating ingoing energy-momentum, as expected.

\begin{figure} 
\centering
\includegraphics[width=4.5in]{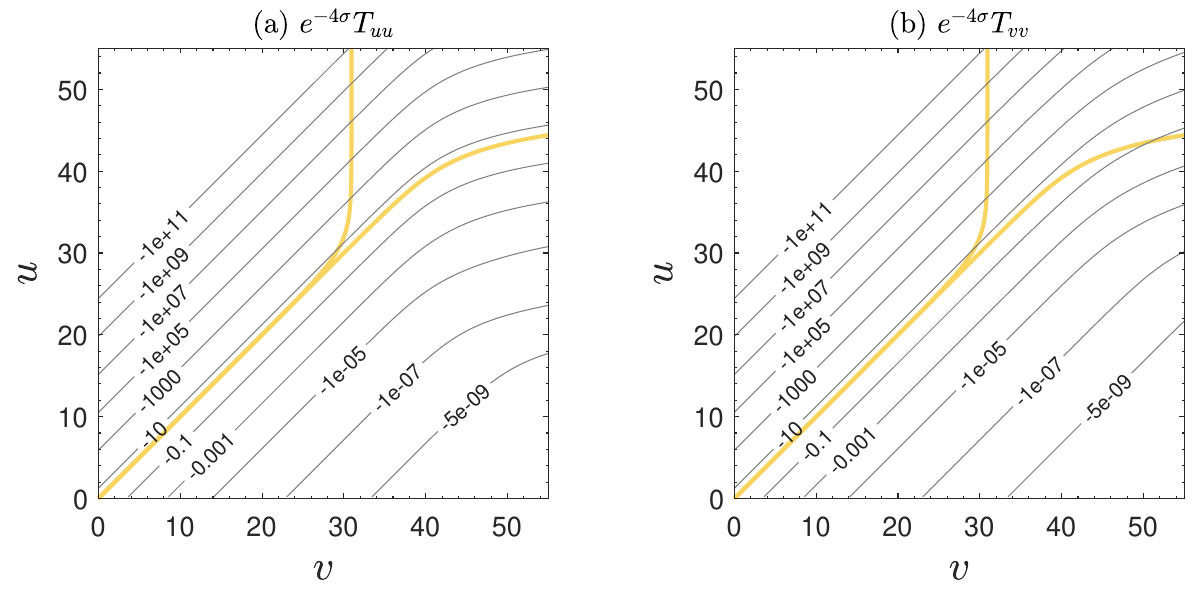} 
\caption{Contour diagrams for the null energy along null geodesics for the same evolution shown in figure \ref{fig:QCSBH r symmetric no pulse}(c) with apparent horizons in yellow.}
\label{fig:QCSBH null energy no pulse}
\end{figure}

Intuitively, we expect the violation of the null energy condition to worsen during wormhole expansion. To get a sense for this, consider a path through the middle of the separated apparent horizons in figures \ref{fig:QCSBH sigma T no pulse}(b) and \ref{fig:QCSBH sigma T no pulse}(c) for $T_{uu}$ and $T_{vv}$. Along this path, we find that $T_{uu}$ and $T_{vv}$ increase, going from initially negative values toward zero. We recall that these quantities do not give the null energy along null geodesics. We show in figures \ref{fig:QCSBH null energy no pulse}(a) and \ref{fig:QCSBH null energy no pulse}(b) $e^{-4\sigma}T_{uu}$ and $e^{-4\sigma} T_{vv}$, which are seen to decrease along the path through the middle of the horizons, as expected.

The results presented in this subsection are for the dynamical evolution of the static solution with $M = 0.5$. We have studied the dynamical evolution of static solutions with other values of $M$ and found them to be qualitatively similar.

%+++++++++++++++++++++++++++++++++++++++++++++++++++++++++

\subsubsection{Wormhole collapse}
\label{sec:QCSBH collapse}

We now introduce an explicit perturbation in the form of a pulse of scalar field, which will cause the wormhole to collapse. For initial data, we use the static solution with $M = 0.5$ augmented by a pulse placed on the $u = u_i$ initial hypersurface with parameters $A = 0.002$, $v_1 = 1$, and $v_2 = 3$. 

\begin{figure} 
\centering
\includegraphics[width=4.5in]{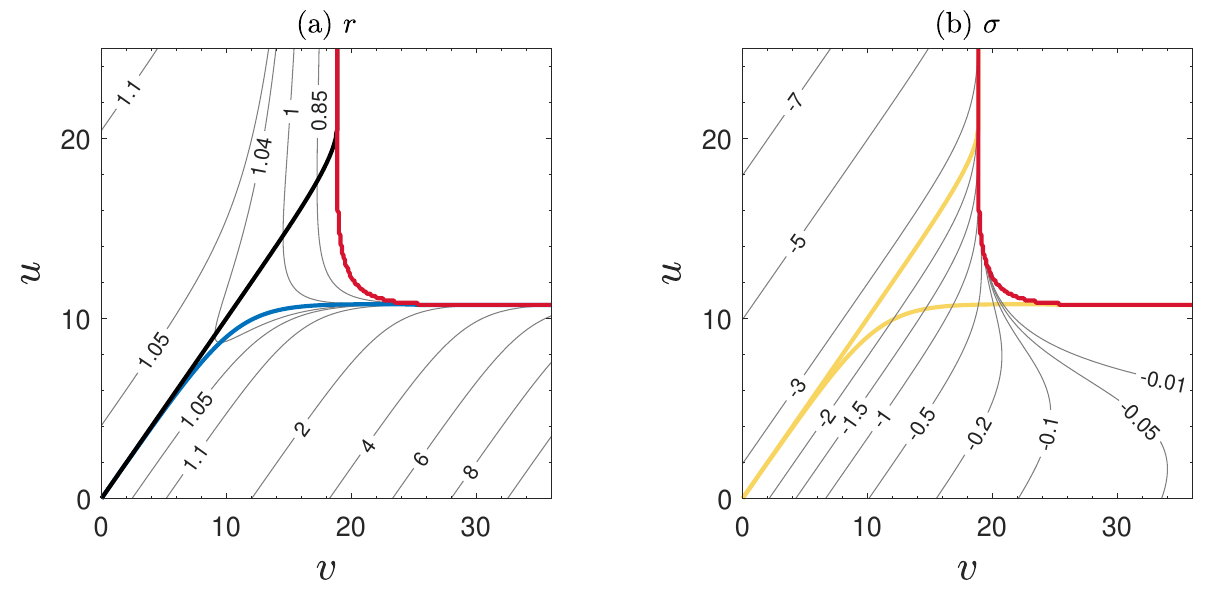} 
\caption{(a) Contour diagram for the areal radius, $r$, for the dynamical evolution of a quantum corrected Schwarzschild black hole with $M = 0.5$ augmented by a scalar field pulse defined by $A = 0.002$, $v_1 = 1$, and $v_2 = 3$. The black curve displays the apparent horizon $r_{,u} = 0$ and the blue curve displays the apparent horizon $r_{,v} = 0$. The red curve displays the central singularity. The region beyond the red curve is not part of the spacetime. The diagram is made using grid spacing $\Delta u = \Delta v = 1/400$. (b) Contour diagram for the metric field $\sigma$ for the same evolution shown in (a), with the apparent horizons shown in yellow.}
\label{fig:QCSBH r sigma with pulse}
\end{figure}

In figure \ref{fig:QCSBH r sigma with pulse}(a), we show the contour diagram for the areal radius. Comparing to the static solution in figure \ref{fig:static QCSBH uv}(b), it is clear that the diagrams are different, indicating that the system in figure \ref{fig:QCSBH r sigma with pulse}(a) evolves away from the static solution. Near the bottom left corner, figures \ref{fig:QCSBH r sigma with pulse}(a) and \ref{fig:static QCSBH uv}(b) look the same, indicating that the system initially maintains the static configuration. The apparent horizons separate at around $u \approx 7$, at which point the system evolves away from the static configuration. In between the separated apparent horizons, moving in a timelike direction moves across decreasing radii and we find that the wormhole is collapsing. Eventually the areal radius collapses to $\sqrt{p}$ and a central singularity forms, which is marked by the red curve. The region beyond the red curve is not part of the spacetime. The black and blue curves evolve to being vertical and horizontal, signifying event horizons at $v\approx 19$ and $u \approx 11$. Black holes have formed which enclose the mouths of the wormhole. We show the contour diagram for the metric field $\sigma$ in figure \ref{fig:QCSBH r sigma with pulse}(b).

As the wormhole collapses, we can see from figure \ref{fig:QCSBH r sigma with pulse}(a) that the apparent horizons evolve along decreasing radii. It was shown in \cite{Kain:2025lxd} that the radius of the $r_{,v} = 0$ apparent horizon (the blue curve) continues to decrease for larger values of $v$, which cannot be determined from figure \ref{fig:QCSBH r sigma with pulse}(a). This was shown by evolving the system using the adaptive gauge method \cite{Eilon:2015axa}, which makes use of different coordinate gauges. There are two possibilities for how the radius of an apparent horizon can decrease for quantum corrected Schwarzschild black holes. First, it is possible because the null energy condition is violated \cite{Gonzalez:2008xk}. This is the most likely explanation during the initial stages of collapse. Second, the renormalized energy-momentum tensor allows for black hole evaporation, which leads to decreasing radii for apparent horizons \cite{Parentani:1994ij, Ayal:1997ab}. This second possibility explains why the radius of the $r_{,v} = 0$ apparent horizon continues to decrease for larger values of $v$.

\begin{figure} 
\centering
\includegraphics[width=6.5in]{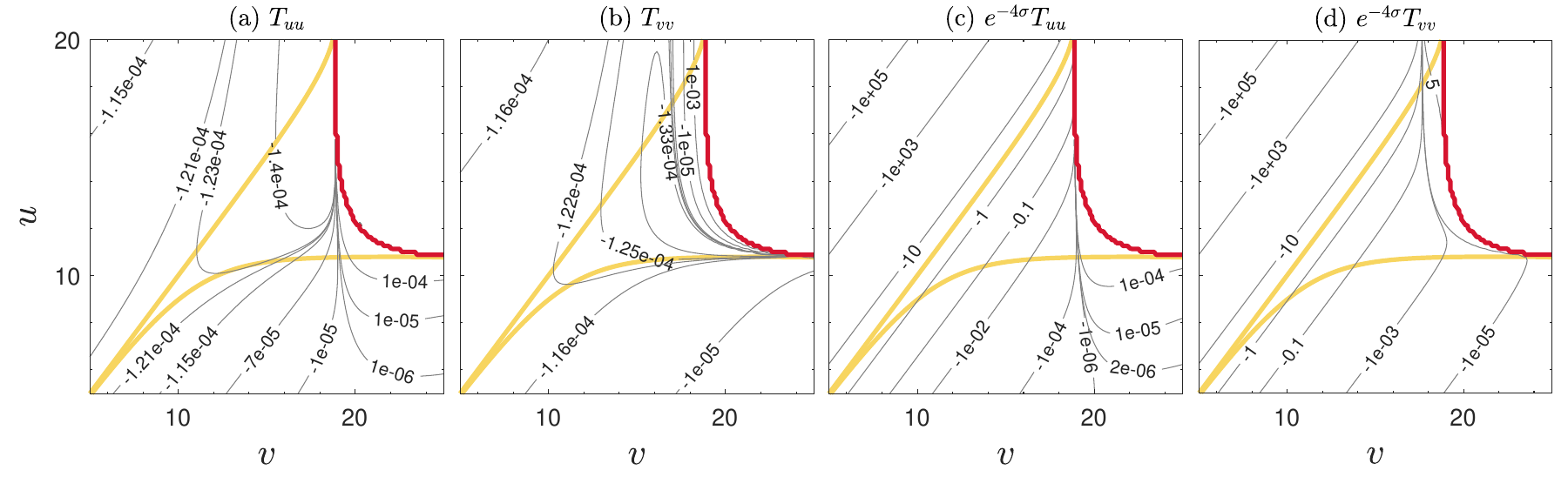} 
\caption{Contour diagrams for (a) outgoing energy-momentum, (b) ingoing energy-momentum, and (c, d) the null energy along null geodesics for the same evolution shown in figure \ref{fig:QCSBH r sigma with pulse}(a), with apparent horizons shown in yellow.}
\label{fig:QCSBH T null energy with pulse}
\end{figure}

The mass inside the apparent horizons decreases during collapse since the radii of the apparent horizons decrease. For the mass to decrease, we expect energy to flow outward. Unlike with the Ellis-Bronnikov wormhole, there is no ghost field and we do not expect energy to also flow inward. We show the contour diagram for outgoing energy-momentum in figure \ref{fig:QCSBH T null energy with pulse}(a). Near future null infinity, $T_{uu}$ is positive, indicating outgoing energy-momentum. We show the contour diagram for ingoing energy-momentum in figure \ref{fig:QCSBH T null energy with pulse}(b). Near future null infinity, $T_{vv}$ is negative, again indicating outgoing energy-momentum. The net flow of energy-momentum near future null infinity is therefore outgoing, as expected.

Moving along a path through the middle of the two separated apparent horizons in figures \ref{fig:QCSBH T null energy with pulse}(a) and \ref{fig:QCSBH T null energy with pulse}(b), we mostly move along decreasing energy-momentum, energy-momentum that is initially negative and becoming more negative. Intuitively, we expect the violation of the null energy condition to weaken as the wormhole collapses. We show contour diagrams for the null energy along null geodesics in figures \ref{fig:QCSBH T null energy with pulse}(c) and \ref{fig:QCSBH T null energy with pulse}(d), where we can see that a path through the middle of the separated horizons moves along increasing energy-momentum, as expected.

We have studied dynamical evolutions with different initial data which also lead to collapse. This includes using different static solutions and different parameters for the pulse. We found the results to be qualitatively similar.

%==========================================================

\section{Comparison and conclusion}
\label{sec:conclusion}

We reviewed the dynamical evolution of spherically symmetric charge free wormholes from an initial static solution in double null coordinates. An advantage of double null coordinates for spherically symmetric simulations is that large potions of spacetime can be presented in a single diagram. Focusing on two specific example systems, we presented results for wormhole expansion and wormhole collapse and studied how properties of these dynamically evolving wormholes could be understood from contour diagrams of the areal radius and energy-momentum tensor.

The first example, which has been extensively studied, was the Ellis-Bronnikov wormhole. This wormhole is sourced by a ghost scalar field. An advantage of this model is its simplicity, since the static solutions can be found analytically, but its dependence on a ghost field makes its physical reasonableness debatable.

The second example was the quantum corrected Schwarzschild black hole in semiclassical gravity, which has a wormhole structure sourced by a renormalized energy-momentum tensor in the Polyakov approximation. Semiclassical gravity and the inclusion of the renormalized energy-momentum tensor may be one of the best motivated extensions to general relativity, making this model much more physically reasonable than the Ellis-Bronnikov wormhole, but it follows from a more complicated set of equations and must be found numerically.

In this final section, we compare the simulations and discuss similarities. Static solutions for both systems are nonlinearly unstable with respect to time dependent perturbations. Determining the stability of static solutions is a common motivation for performing a dynamical evolution in the first place. The dynamical evolution has the added benefit of also determining the configuration to which an unstable static solution will evolve. 

In the absence of an explicit perturbation, both systems exhibit wormhole expansion. The diagrams for the areal radius for symmetric and asymmetric Ellis-Bronnikov wormholes shown in figures \ref{fig:EB r symmetric no pulse} and \ref{fig:EB r asymmetric no pulse} look similar to the analogous diagrams for the quantum corrected Schwarzschild black hole shown in figure \ref{fig:QCSBH r symmetric no pulse}. In both systems, as the wormhole expands, the radii of the apparent horizons and the Misner-Sharp mass contained inside the apparent horizons increases. Consistent with this mass increase, the net flow of energy-momentum is ingoing.

$T_{uu}$ and $T_{vv}$ are often used for the null energy and it is true that the null energy condition is violated if either one is negative. We stressed that it is $e^{-4\sigma}T_{uu}$ and $e^{-4\sigma}T_{vv}$ that gives the null energy along null geodesics and that it is these quantities that are referenced in the null energy condition as derived from the Raychaudhuri equation. We showed that the null energy along null geodesics generally decreases in both systems, worsening the violation of the null energy condition as the wormhole expands, which is inline with expectations, while $T_{uu}$ and $T_{vv}$ do not necessarily decrease.

An explicit perturbation in the form of a pulse of (regular) scalar field leads to wormhole collapse in both systems. The wormhole throat radius decreases, closing off the ability to travel from one side of the wormhole to the other. As the wormhole collapses, black holes form, which enclose the mouths of the wormhole, and a curvature singularity forms. The contour diagrams for the areal radius for the Ellis-Bronnikov wormhole in figures \ref{fig:EB r symmetric with pulse}, \ref{fig:EB r asymmetric with pulse}(a), and \ref{fig:EB r asymmetric with pulse}(b) have obvious similarities to the analogous diagram for the quantum corrected Schwarzschild black hole in figure \ref{fig:QCSBH r sigma with pulse}. 

The collapsed Ellis-Bronnikov wormhole asymptotically approaches a static Schwarzschild black hole on both sides. The collapsed quantum corrected Schwarzschild black hole cannot approach a static black hole, because the renormalized energy-momentum tensor allows for black hole evaporation and an evaporating black hole is not static. It was shown in \cite{Kain:2025lxd} that the side of the collapsed wormhole with larger values of $v$ approaches an evaporating Schwarzschild black hole, while the black hole on the other side of the collapsed wormhole has not yet been sufficiently studied.

In both systems, as the wormhole collapses, the radii of the apparent horizons and the Miser-Sharp mass contained inside the apparent horizons decreases. Consistent with this decrease in mass, we found for both systems portions of spacetime in which the net flow of energy-momentum is outgoing. The Ellis-Bronnikov wormhole is more complicated because it must expel the ghost field in order to settle down to a Schwarzschild black hole. This requires it to also have a net flow of outgoing negative energy or, equivalently, ingoing energy-momentum. 

The null energy along null geodesics, $e^{-4\sigma}T_{uu}$ and $e^{-4\sigma}T_{vv}$, generally increases for both systems and the violation of the null energy condition weakens, as the wormhole collapses, which is in line with expectations, while $T_{uu}$ and $T_{vv}$ do not necessarily increase.

As far as we are aware, the Ellis-Bronnikov wormhole and the quantum corrected Schwarzschild black hole are the only two charge free wormhole systems that have been dynamically evolved and which exhibit expansion or collapse. Given that their respective wormhole structures follow from very different sources---a ghost field on the one hand and a renormalized energy-momentum tensor on the other---yet their dynamical results are remarkably similar, we speculate that their shared properties are generic. Consequently, the Ellis-Bronnikov wormhole, with its relatively simple analytic static solutions, is a convenient and useful model for understanding dynamical wormhole expansion and collapse in general, notwithstanding its use of a ghost field.

%====================================================================

\appendix
\renewcommand\thesection{\Alph{section}}

\section{Code tests}
\label{app:code tests}

We have written code to numerically evolve wormhole systems. It is necessary to test the code for convergence. This includes testing for convergence to the intended second order accuracy and for convergence to a consistent solution. In this appendix, we explain how this can be done and present some tests of our code.

We can test for second order convergence as follows. When performing a dynamical evolution, we use grid spacing $\Delta u = \Delta v= 1/ N$, where $N$ is the number of grid points per unit interval. We perform three evolutions, where each evolution uses the same initial data, but uses a different grid spacing. We then compute the convergence functions
\begin{eqnarray} \label{convergence functions}
C_f(N_1,N_2) &= \sum_i | f_i(N_1) - f_i(N_2)|
\nonumber \\
C_f(N_2,N_3) &= \sum_i | f_i(N_2) - f_i(N_3)|,
\end{eqnarray}
where $f_i(N)$ is the value of field $f$ at grid point $i$ computed using $N$ grid points per unit interval. Each of the $f_i$ in (\ref{convergence functions}) must be computed at the \textit{same} grid point. The sum may be taken over grid points along a row with constant $u$ or a column with constant $v$. If our results are second order convergent, then if the grid spacings $\Delta u$ and $\Delta v$ decrease by a factor of 2, so that $N_3 = 2N_2 = 4N_1$, the convergence functions will decrease by a factor of 4, so that $C_f(N_1, N_2) = 4C_f(N_2,N_3)$ \cite{AlcubierreBook}.

We cannot apply this test to an unstable static solution without including an explicit perturbation. To understand why, consider figure \ref{fig:EB r symmetric no pulse} for the expansion of the symmetric massless Ellis-Bronnikov wormhole. Figure \ref{fig:EB r symmetric no pulse} displays results at three different grid spacings, where we find that the three evolutions are different. Indeed, the fact that they are different indicates that the static solution is unstable. If we were to evaluate the convergence functions in (\ref{convergence functions}) using these results, we would be comparing different evolutions, instead of comparing evolutions that are converging as $N$ is increased.

\begin{figure} 
\centering
\includegraphics[width=4.5in]{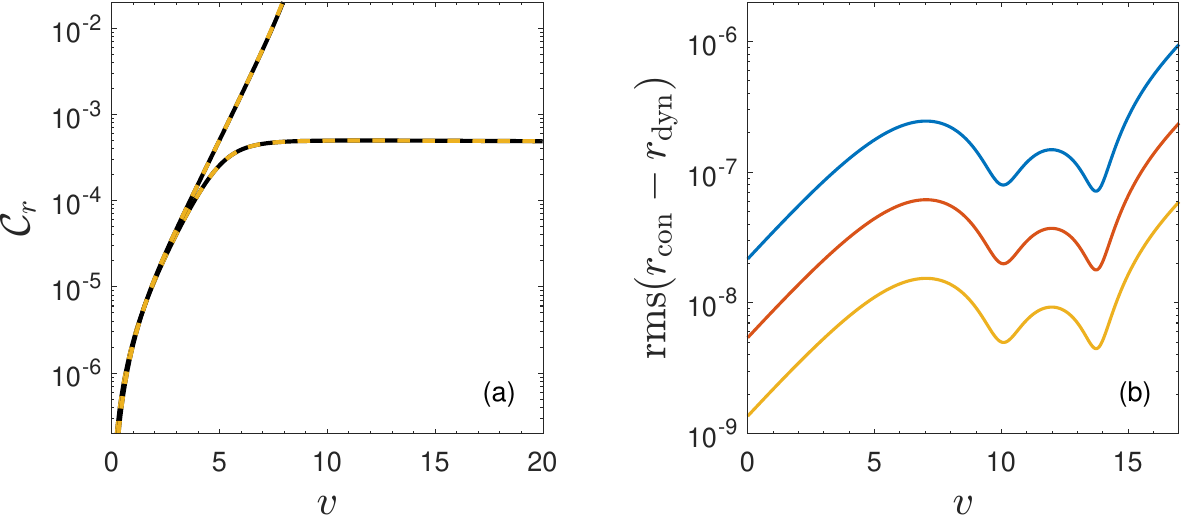} 
\caption{(a) Convergence functions in (\ref{convergence functions}) evaluated for three evolutions using the same initial data as figure \ref{fig:EB r symmetric with pulse} using $N_1 = 100$, $N_2 = 200$, and $N_3 = 400$. The solid black curves plot $C_r(100, 200)$ and the dashed yellow curves plot $4C_r(200, 400)$. The lower curve that flattens out is for $u = 5$ and the upper curve is for $u = 17$. That the curves overlap indicates second order convergence. (b) Equation (\ref{rms eval}) is plotted for three evolutions using the same initial data as figure \ref{fig:QCSBH r sigma with pulse} and grid spacing $\Delta u = \Delta v = 1/N$ with $N = 100$ (blue), 200 (red), and 400 (yellow) grid points per interval. That each curve drops by a factor of 4 from the previous curve indicates second order convergence. That the curves have small values indicates that the constraint equation is satisfied.}
\label{fig:convergence 1}
\end{figure}

We can apply this test to evolutions which include a scalar field pulse. As an example, consider three evolutions using the same initial data used in figure \ref{fig:EB r symmetric with pulse}, which shows the collapse of the symmetric massless Ellis-Bronnikov wormhole, and using $N_1 = 100$, $N_2 = 200$, and $N_3 = 400$. In figure \ref{fig:convergence 1}(a), we show $C_r(100, 200)$ as the black curve and $4C_r(200, 400)$ as the dashed yellow curve. Figure \ref{fig:convergence 1}(a) performs the sum in the convergence functions over rows of constant $u$. The lower curve that flattens out is for $u = 5$ and the upper curve is for $u = 17$. We compute a set of convergence functions for all values of $v$. We find that the dashed yellow curves agree with the solid black curves, indicating second order convergence. The upper curve veers upward because the curve eventually runs into the singularity. 

We can also test whether our results are converging to a consistent solution. Our dynamical evolution code primarily makes use of the evolution equations in (\ref{sigma r evo}), but not the constraint equations in (\ref{constraint eqs}). As such, the constraint equations are available for code testing. The solution to either constraint equation is $r$. If we solve one of the constraint equations, we can compare the solution for $r$ to the value computed from our code. We solve the constraint equation using second order Runge-Kutta. For the non-$r$ fields in the constraint equation we use the results from our code and for the derivatives we use standard second order finite difference formulas. At each grid point, we can compute the difference between the value of $r$ from the dynamical evolution and the value of $r$ from the constraint equation. If we'd like to compare results which use different grid spacings, then we will need to make sure that we compare results at the \textit{same} grid points. A less tedious alternative is to compute the difference at \textit{every} grid point and then compute the root-mean-square (rms) value,
\begin{equation} \label{rms eval}
\mathrm{rms}(r_\mathrm{con} - r_\mathrm{dyn}).
\end{equation}
We can then compare rms values which use different grid spacings.
The rms value is to be computed along columns of constant $v$ when using the top constraint equation in (\ref{constraint eqs}) and along rows of constant $u$ when using the bottom constraint equation in (\ref{constraint eqs}).

As an example for this test, we use the same initial data used in figure \ref{fig:QCSBH r sigma with pulse}, which shows the collapse of a quantum corrected Schwarzschild black hole, and evolve the system three times using $N = 100$, 200, and 400 grid points per unit interval. For each evolution, we compute (\ref{rms eval}) along columns of constant $v$ and use the top constraint equation in (\ref{constraint eqs with RSET}), which is specific to the quantum corrected Schwarzschild black hole. We do this for a range of $v$ values. The result is shown in figure \ref{fig:convergence 1}(b). Note that we stop plotting the curves before they get too close to the singularity, which happens at around $v\approx 19$. There are two main takeaways from this figure. First, just as with figure \ref{fig:convergence 1}(a), we find that when the grid spacing drops by a factor of 2, the curves drop by a factor of 4, which indicates second order convergence. Second, the results are small and getting smaller as we decrease the grid spacing, which indicates that the constraint equation is satisfied and that the evolution is converging to a consistent solution.

%====================================================================

%\begin{thebibliography}{99}
%\end{thebibliography}
% for bibtex
%\bibliographystyle{iopart-num}
%\bibliographystyle{apsrev4-2}
%\bibliographystyle{IEEEtran}
%\bibliography{wormholes}

% for biblatex
%\printbibliography

%=======================================================================


\end{document}